\documentclass[preprint]{JHEP3}  
\usepackage{graphicx}
\usepackage{bm}
\usepackage{dcolumn}
\usepackage{amsmath}
\usepackage{amsfonts}
\usepackage{amssymb}

\title{Semileptonic hyperon decays in the self-consistent SU(3) chiral 
quark-soliton model}

\author{Tim Ledwig \\ Institut f\"ur Theoretische Physik II,
  Ruhr-Universit\" at Bochum, D--44780 Bochum, Germany \\
E-mail: \email{Tim.Ledwig@tp2.rub.de}
}
\author{Antonio Silva \\
Centro de Fisica Computacional (CFC), Departamento de
  Fisica, Universidade de Coimbra, P-3004-516 Coimbra, Portugal \\
Faculdade de Engenharia da Universidade do Porto,
  P-4200-465 Porto, Portugal\\
E-mail: \email{ajsilva@fe.up.pt}
}
\author{Hyun-Chul Kim \\ Department of Physics, Inha University,
  Incheon 402-751, Republic of Korea \\
E-mail: \email{hchkim@inha.ac.kr}
}
\author{Klaus Goeke \\ Institut f\"ur Theoretische Physik II,
  Ruhr-Universit\" at Bochum, D--44780 Bochum, Germany \\
ECT*, Villa Tambosi, Strada delle Tabarelle 286, I-38050
  Villazzano (TN), Italy
E-mail: \email{Klaus.Goeke@tp2.rub.de}
}

\abstract{
We investigate the semileptonic hyperon decays within the framework
of the self-consistent SU(3) chiral quark-soliton model ($\chi$QSM).
We take linear $1/N_c$ rotational as well as linear $m_s$ corrections
into account and apply the symmetry conserving quantization.  We
present the results for the form factors $f_1(Q^2)$, $f_2(Q^2)$ and
$g_1(Q^2)$ in addition to the semileptonic decay constants of
hyperons.  We also have calculated the radii and dipole masses of
these form factors for all relevant strangeness-conserving and
strangeness-changing transitions. }


\preprint{INHA-NTG-10/2008}

\begin{document}

\section{Introduction}
In the Standard model, the Cabibbo-Kobayashi-Maskawa matrix elements
$|V_{ud}|$ and $|V_{us}|$ characterize the transition amplitudes
for light hadron processes involving the quark transitions $d\to
ue^{-}\overline{\nu}_{e}$ and $s\to
ue^{-}\overline{\nu}_{e}$~\cite{Cabibbo1963,CKM1973}. At present, the
semileptonic kaon and $\pi^{+}$ decays~\cite{Status_of_Cabibbo_angle,
V(us)fromK(l3)} provide the most precise measurement of the product
$|V_{us}\cdot f_{+}(0)|$ with $f_{+}(0)$ being the kaon vector form
factor.  Generally, the Ademollo-Gatto theorem~\cite{AG_theorem}
states that the linear-order flavor $SU(3)$ breaking effects vanish in
the vector matrix elements between hadron states in the same
multiplet. This enables us to extract $V_{us}$ from 
kaon decays with high accuracy.  On the other hand, the semileptonic
hyperon decays (SHD) can be also used for an independent determination
of $V_{us}$ and $V_{ud}$~\cite{Cabibbo2003,Cabibbo2004}, where the
product of $|V_{us}\cdot f_{1}(0)|$, $f_{1}(0)$ being the hyperon
vector form factor, is accessed. Reference~\cite{Cabibbo2003} has
suggested to use recent experiments to compare the $V_{us}$ from
the kaon with the SHD data.

Motivated by a recent progress in this field, we review in the present
work the investigation of the semileptonic hyperon form factors in
the chiral quark-soliton model ($\chi$QSM). Since the first results
from the $\chi$QSM~\cite{CQSM_semilep1998} on this issue, several new 
experimental and theoretical works have been developed.  The KTeV
collaboration first reported the measurement of the $\Xi^{0} \to
\Sigma^{+} e^{-} \overline{\nu}_{e}$ decay~\cite{KTeV1999}
followed by the first determination of these form factors
\cite{KTeV2001}.  They published for this process the ratios
$f_{2}(0)/f_{1}(0)=2.0\pm1.2_{\textrm{stat}}\pm0.5_{\textrm{syst}}$
and $g_{1}(0)/f_{1}(0)=1.32_{-0.17\textrm{stat}}^{+0.21}\pm0.05_{\textrm{syst}}$.
Very recently, the NA48/1 collaboration reported branching ratios for
the same process with improved statistics, compared to the KTeV
experiments, and announced $g_{1}(0)/f_{1}(0)=1.20\pm0.05$ for
$\Xi^{0} \to \Sigma^{+}e^{-} \overline{\nu}_{e}$~\cite{NA48/I:2007}.
This particular SHD are interesting, since in flavor $SU(3)$
symmetry the $g_1/f_1$ ratio is equal to that of the neutron $\beta$
decay.  The isospin partner process $\Xi^{-} \to \Sigma^{0} e^{-}
\overline{\nu}$ was investigated in Ref.~\cite{BGHORS:1983}, where the
ratio $g_{1}(0) / f_{1}(0) = 1.25_{-0.16}^{+0.14}$ is presented
together with the results for $\Lambda\to pe^{-}\overline{\nu}$
and $\Xi^{-} \to \Lambda e^{-}
\overline{\nu}$. References~\cite{BGHORS:1982,LamPro} invstigated
$\Sigma^{-} \to \Lambda e^{-} \overline{\nu}$ as well as $\Lambda \to
pe^{-} \overline{\nu}$.

On the theoretical side, chiral perturbation theory ($\chi$PT) is used
to gain information about flavor $SU(3)$ breaking effects to $f_1(0)$
beyond the first order~\cite{HyperonVFF,HyperonCHPT_Meissner,
ChiralVecCorr,ChiralVecExtrap}, where other form factors of
SHD have been also investigated. The results for $V_{us}$ and SHD data
were discussed in Ref.~\cite{HyperonV(us)1} and for the large
$N_{c}$ expansion in Ref.~\cite{HyperonV(us)2}.  Very recently,
Ref.~\cite{LatticeSigmaNeutron} has also reported the first quenched
lattice QCD study for all form factors of the $\Sigma^{-}\to nl\nu$
decay, in which the results were found to be:
$f_{2}(0)/f_{1}(0)=-1.52\pm0.81$ and $g_{1}(0) / f_{1}(0) = -0.287 \pm
0.052$.  In addition, Ref.\cite{BetaDecayRelQuarkModel} presents a
study of SHD in a relativistic constituent quark model.  We will see
that our results are in qualitative agreement with these works as well
as with the above mentioned experimental data.

In the present work we will investigate the vector $f_{1}(Q^{2})$,
$f_{2}(Q^{2})$ and axial-vector $g_{1}(Q^{2})$ form factors for
strangeness-conserving and strangeness-changing SHD in the
self-consistent SU(3) $\chi$QSM. Since the form factors $f_{3}(Q^{2})$
and $g_{3}(Q^{2})$ are always multiplied by a factor of $m_{e}/M_{B}$
in the transition amplitude, they can be safely neglected.  The value
of the form factor $g_{2}(Q^{2})$ at $Q^{2}=0$ vanishes in exact SU(3)
symmetry.  Reference~\cite{CQSM_semilep1998} presented in 1998 the SHD
constants $f_{1}(0)$, $f_{2}(0)$ and $g_{1}(0)$ within the same
framework of the $\chi$QSM. However, since then new experimental data
as well as new theoretical results, some mentioned above, have been
progressed and, moreover, Ref.~\cite{Praszalowicz:1998jm} proposed in
the $\chi$QSM the symmetry-conserving quantization that makes the
Gell-Mann-Nishijima relation well satisfied.  We extend, therefore, in
this work the previous investigation to the full form factors up to
$Q^{2}\leq1\textrm{GeV}^{2}$ with the symmetry-conserving quantization
employed~\cite{Praszalowicz:1998jm}.  We take into account linear
$1/N_c$ rotational corrections as well as linear $m_s$ corrections.

A merit of the $\chi$QSM lies in the fact that we are able to
calculate various unrelated baryonic observables within the same
setting, once having determined the eigenvalues of the $\chi$QSM
one-particle Dirac Hamiltonian.  These eigenvalues were obtained by
diagonalizing the Hamiltonian numerically with a self-consistent meson
profile of the soliton, which is found in an action principle of
minimizing the nucleon mass in the $\chi$QSM.  The same eigenvalues
were used in the past for investigating the mass splittings of the
SU(3) baryons, form factors, and parton- and
antiparton-distributions~\cite{Christov:1995vm,Dressler:2001,
GoekePPSU:2001,SchweitzerUPWPG:2001,OssmannPSUG:2005,SilvaKUG:2005,
WakamatsuN:2006,Wakamatsu:2005,Wakamatsuref:2003,Wakamatsu:2000,
Goeke:2006gi,Silva:2005qm}, all relevant sum rules of these
observables being  simultaneously fulfilled.  Nearly all observables
are in good agreement with experimental data showing errors between
$(10 \sim 30)\,\%$.  In particular, the dependence of all form factors
on the momentum transfer is well reproduced within the
$\chi$QSM. References~\cite{Silva:2005qm,GoekeKSU2007} have used the same
set of fixed parameters as in the present work to investigate the
strange electromagnetic form factors and the parity-violating
asymmetries of polarized electron-proton scattering. In particular,
the six electromagnetic form factors $G_{E,M}^{u,d,s}(Q^2)$ and three
axial-vector form factors $G_A^{u,d,s}(Q^2)$ were calculated and are
all in good agreement with experimental data.  The vector and
axial-vector form factors used in these $\chi$QSM-works are the
basis for the present investigation.  In addition, the same techniques were
recently used for calculating observables of the anti-decuplet
pentaquarks~\cite{vector_Theta,decay_Theta}. The results describe well
the experimental data avalilable so far, though their relevance is
controversially discussed. 

The present work is sketched as follows. In Section II, we
present the general formalism for the SHD.  In Section III, we
show how to compute the observables of SHD within the selfconsistent
SU(3) $\chi$QSM.  In Section IV, we present and discuss the obtained
results.  In the final Section, we summarize the present work and draw
conclusions.  Additional detailed formulae are given in Appendices.
\section{General Formalism}
The decay rates for the SHD processes $B_{1}\to
B_{2}l\overline{\nu}_{l}$ are determined by the transition matrix
element $\mathcal{M}_{B_{1}\to B_{2}l\overline{\nu}_{l}}$ that is
generally written as
\begin{equation}
\mathcal{M}_{B_{1}\to B_{2}l\overline{\nu}_{l}}=\frac{G_F}{\sqrt{2}}\,
V\,\langle
B_{2}|J_{W}^{\mu}|B_{1}\rangle\overline{u}_{l}(p_{l})\gamma_{\mu}(1 -
\gamma^{5})u_{\overline{\nu}_{l}}(p_{\nu}),
\end{equation}
where $G_F$ denotes the well-known Fermi coupling constant and
$V=V_{ud}(V_{us})$ stand for the Cabibbo-Kobayashi-Maskawa angles for
$\Delta S=0$ ($\Delta S=1$) processes, respectively. It is the
hadronic matrix element $\langle B_{2}|J_{W}^{\mu}|B_{1}\rangle$
that will be considered in this work for the $\Delta S=0$ transitions
within the baryon octet:
\begin{equation}
n\to p,\,\,\,\,\,\,\,\,\,\,\Sigma^{-}\to\Lambda,\,\,\,\,\,\,\,\,\,\,
\Sigma^{-}\to\Sigma^{0},\,\,\,\,\,\,\,\,\,\,\Xi^{-}\to\Xi^{0},\,\,\,
\label{eq:DS0 processes}
\end{equation}
and $\Delta S=1$ transitions
\begin{equation}
\Sigma^{-}\to n,\,\,\,\,\,\,\,\,\,\,\Lambda\to
p,\,\,\,\,\,\,\,\,\,\,\Xi^{-}\to\Sigma^{0},\,\,\,\,\,\,\,\,\,\,
\Xi^{-}\to\Lambda.
\label{eq:DS1 processes}
\end{equation}
All other transitions are related to these ones by isospin symmetry,
of which the relations are given in the Appendix.  The transition
matrix elements for these two types of transitions are written
explicitly as 
\begin{eqnarray}
\langle B_{2}(p_{2})|J_{W}^{\mu}(0)|B_{1}(p_{1})\rangle & = & \langle
B_{2}(p_{2})|\overline{\psi}(0)\gamma^{\mu}(1-\gamma^{5})\frac{1}{2}
\Big(\lambda^{1(4)}\pm
i\lambda^{2(5)}\Big)\psi(0)|B_{1}(p_{1})\rangle,
\end{eqnarray}
where $\lambda^{\chi=1,2,4,5}$ are flavor SU(3) Gell-Mann matrices.
Thus, we need to consider the following matrix elements for the vector
current:
\begin{eqnarray}
\langle B_{2}|J_{V}^{\mu\chi}|B_{1}\rangle & = &
\langle B_{2}(p_{2})|\overline{\psi}(0)\gamma^{\mu}
\frac{\lambda^{\chi}}{2} \psi(0)|B_{1}(p_{1})\rangle \cr
 & = &  \overline{u}_{2}(p_{2})\Big[ f_{1}^{B_{1}B_{2}}(Q^{2})
 \gamma^{\mu} + \frac{f_{2}^{B_{1}B_{2}}(Q^{2})i
   \sigma^{\mu\beta}q_{\beta}}{M_{B_1}} +
 \frac{f_{3}^{B_{1}B_{2}}(Q^{2})q^{\mu}}{M_{B_1}} \Big]
 \frac{\lambda^{\chi}}{2} u_{1}(p_{1}),
\label{eq:vector-general-current}
\end{eqnarray}
and for the axial-vector current:
\begin{eqnarray}
\langle B_{2}|J_{A}^{\mu\chi}|B_{1}\rangle & = &
\langle B_{2}(p_{2})|\overline{\psi}(0)\gamma^{\mu}\gamma^{5}
\frac{\lambda^{\chi}}{2}\psi(0)|B_{1}(p_{1})\rangle\cr
 & = &
 \overline{u}_{2}(p_{2})\Big[g_{1}^{B_{1}B_{2}}(Q^{2})
\gamma^{\mu}+\frac{g_{2}^{B_{1}B_{2}}(Q^{2})i\sigma^{\mu\beta}q_{\beta}}{M_{B_1}} 
\cr
&& \hspace{6cm}
+\;\frac{g_{3}^{B_{1}B_{2}}(Q^{2})q^{\mu}}{M_{B_1}}\Big]\gamma^{5}
\frac{\lambda^{\chi}}{2}u_{1}(p_{1}),
\label{eq:axial-general-current}
\end{eqnarray}
where the form factors are normalized to the mass $M_{B_1}$ of the
decaying particle. The form factors are known to be the real functions
of the momentum-transfer $Q^{2}=-q^{2}$ with $q=p_{2}-p_{1}$.  We
concentrate on the form factors $f_{1}^{B_{1}B_{2}}(Q^{2})$,
$f_{2}^{B_{1}B_{2}}(Q^{2})$ and $g_{1}^{B_{1}B_{2}}(Q^{2})$.  The
$f_{3}^{B_{1}B_{2}}(Q^{2})$, $g_{3}^{B_{1}B_{2}}(Q^{2})$ and
$g_{2}^{B_{1}B_{2}}(Q^{2})$ are neglected since the first two are
always suppressed by a factor of $(m_e/M_B)^2$ due to $q^\mu$ in the
cross section and the latter one is entirely due to the SU(3) symmetry
breaking that is small for the octet baryons. We will first determine
the SHD constants at $Q^{2}=0$ and the radii of these form factors for
the transitions mentioned above.  In order to calculate these form 
factors, we will use the rest frame of the decaying baryon $B_{1}$, 
i.e. $p_{1}=(M_{B_1},0)$, $p_{2}=(E_{2},{\bm q})$, where we have the
kinematics
\begin{equation}
{\bm q}^{2}=\Big(\frac{Q^{2}+M_{B_1}^{2}+M_{B_2}^{2}}{2M_{B_1}}\Big)^{2} -
M_{B_2}^{2},\,\,\,\,\,\,\,\,\,\,\,\,
E_{2}=\frac{Q^{2}+M_{B_2}^{2}+M_{B_1}^{2}}{2M_{B_1}}.
\label{vecqdef}
\end{equation}
For the matrix element of the vector current given in
Eq.(\ref{eq:vector-general-current}), it is more convenient in the
present scheme to calculate the Sachs-type
form factors $G_{E}^{B_{1}B_{2}}(Q^{2})$ and
$G_{M}^{B_{1}B_{2}}(Q^{2})$, since they are directly related to the
matrix elements of the time and space components of the vector
current as follows:
\begin{eqnarray}
G_{E}^{B_{1}B_{2}}(Q^{2}) &=& \int\frac{d\Omega_{q}}{4\pi} \langle
B_{2}(p_{2})|J_{V}^{0}(0)|B_{1}(p_{1})\rangle,\label{GE}\\
G_{M}^{B_{1}B_{2}}(Q^{2}) &=& 3M_{p}\int\frac{d\Omega_{q}}{4\pi}
\frac{q^{i}\epsilon^{ik3}}{i\mid{\bm q}\mid^{2}}\langle
B_{2}(p_{2})|J_{V}^{k}(0)|B_{1}(p_{1})\rangle,
\label{GM}
\end{eqnarray}
where  $G_{M}^{B_{1}B_{2}}(Q^{2})$ is multiplied by the proton mass
$M_{p}$.  This gives in the rest frame of $B_{1}$ by explicitly
applying those projections to the right-hand side of
Eq.(\ref{eq:vector-general-current}) for equal initial and final
baryon spins:
\begin{eqnarray}
G_{E}^{B_{1}B_{2}}(Q^{2}) & = & Nf_{1}^{B_{1}B_{2}}(Q^{2}) -
f_{2}^{B_{1}B_{2}}(Q^{2})\frac{1}{M_{B_1}}\frac{N{\bm q}^{2}}{
E_{2}+M_{B_2}}+f_{3}^{B_{1}B_{2}}(Q^{2})\frac{q^{0}}{M_{B_1}}N,
\label{eq:eff Ge}\\
G_{M}^{B_{1}B_{2}}(Q^{2}) & = &
\frac{2M_{p}N}{E_{2}+M_{B_2}}\Big[f_{1}^{B_{1}B_{2}}(Q^{2}) +
f_{2}^{B_{1}B_{2}}(Q^{2})\frac{M_{B_1}+M_{B_2}}{M_{B_1}}\Big]
\label{eq:eff Gm}
\end{eqnarray}
with $N=\sqrt{(E_{2}+M_{B_2})/(2M_{B_2})}$. The $M_{B_1}$ in the
denominator comes from the normalization of $f_{2}^{B_{1}B_{2}}(Q^{2})$ and
$f_{3}^{B_{1}B_{2}}(Q^{2})$. The operators applied on the time and
space components of the vector current in Eqs.(\ref{GE},\ref{GM}) are
the same that project out the normal Sachs form factors
$G_{E}(Q^{2})$, $G_{M}(Q^{2})$ in the case of, e.g., the proton
electric and magnetic form factors.

In order to extract the form factors $g_{1}^{B_{1}B_{2}}(Q^{2})$, it
is helpful to contract the spacial component of the current in
Eq.(\ref{eq:axial-general-current}), multiplying by ${\bm
  q}\times({\bm q}\times$, and to perform afterwards the average over
the  orientation of the angular momentum. Choosing
the initial and final baryon-spins to be equal, we extract
$g_{1}^{B_{1}B_{2}}(Q^{2})$ from the third spacial component.
\begin{equation}
\int\frac{d\Omega_{q}}{4\pi}{\bm q}\times\Big({\bm q}\times\langle
B_{2}(p_{2})|{\bm J}_{A}(0)|B_{1}(p_{1})\rangle\Big)  =
Ng_{1}^{B_{1}B_{2}}(Q^{2})(-)\frac{2}{3}\,{\bm q}^{2}
\phi_{s^{2}}^{\dagger}{\bm \sigma}\phi_{s^1},
\end{equation}
\begin{equation}
g_{1}^{B_{1}B_{2}}(Q^{2}) = \frac{1}{N}(-)\frac{3}{2}\frac{1}{{\bm
    q}^{2}}\int\frac{d\Omega_{q}}{4\pi}{\bm q}\times\Big({\bm
  q}\times\langle
B_{2}(p_{2})|{\bm J}_{A}(0)|B_{1}(p_{1})\rangle\Big)_{z}.
\label{GA}
\end{equation}
\section{The Form factors in the Chiral Quark-Soliton Model}
In this Section, we show briefly how to compute
Eqs.(\ref{GE},\ref{GM},\ref{GA}) in the SU(3) $\chi$QSM.  For details
we refer to Ref.~\cite{Christov:1995vm,Christov:eleff,Kim:eleff,
Meissner:axialff}. In general, the baryonic matrix elements are
expressed by 
\begin{equation}
\langle B_{2}(p_{2})|\mathcal{J}^{\mu \chi}(0)|B_{1}(p_{1})\rangle =
\langle B_{2}(p_{2})|\overline{\psi}(0) \mathcal{O}^{\mu \chi} \psi(0) 
|B_{1}(p_{1})\rangle,
\label{matelem}
\end{equation}
where the explicit form of $\mathcal{O}^{\mu \chi}$ are found in
Eqs.(\ref{GE},\ref{GM},\ref{GA}). In order to evaluate the matrix
elements in the model, we start from the low-energy effective
partition function defined as follows:
\begin{equation}
\label{eq:part}
\mathcal{Z}_{\mathrm{\chi QSM}} = \int
\mathcal{D}\psi\mathcal{D}\psi^\dagger
 \mathcal{D}U \exp\left[-\int d^4 x\psi^\dagger i D(U)\psi\right]
= \int \mathcal{D} U \exp(-S_{\mathrm{eff}}[U]),
\end{equation}
where $\psi$ and $U$ denote the quark and pseudo-Goldstone boson
fields, respectively.  The $S_{\mathrm{eff}}$ stands for the effective 
chiral action expressed as
\begin{equation}
  \label{eq:echl}
S_{\mathrm{eff}} (U) = -N_c\mathrm{Tr}\ln iD(U),
\end{equation}
where $\mathrm{Tr}$ represents the functional trace, $N_c$ the number
of colors, and $D$ the Dirac differential operator in Euclidean space:
\begin{equation}
 \label{eq:Dirac}
D(U) = \gamma^4(i\rlap{/}{\partial} -\hat{m} -MU^{\gamma_5}) =
-i\partial_4 + h(U) - \delta m.
\end{equation}
We assume isospin symmetry and decompose the current quark mass matrix
into $\hat{m} = \mathrm{diag}(\overline{m}, \,\overline{m},
\,m_{\mathrm{s}}) = \overline{m} + \delta m$ with $\overline{m}$ and
$m_s$ as the average of the up- and down-quark masses and strange
quark mass, respectively.  The $\delta m$ is given by
\begin{equation}
  \label{eq:deltam}
\delta m = M_1\gamma^4\bm 1 + M_8 \gamma^4\lambda^8,
\end{equation}
where $M_1$ and $M_8$ are singlet and octet components of the current
quark masses defined as $M_1 =(-\overline{m}+m_{\mathrm{s}})/3$ and
$M_8=(\overline{m}-m_{\mathrm{s}})/\sqrt{3}$.
The $\partial_4$ designates the derivative
with respect to the Euclidean time and $h(U)$ stands for the Dirac
single-quark Hamiltonian:
\begin{equation}
h(U)= i \gamma^4 \gamma^i\partial_i - \gamma^4 MU^{\gamma_5}
-\gamma^4\overline{m}. 
\label{eq:diracham}
\end{equation}
For the chiral field $U^{\gamma_5}$, we assume Witten's embedding of
the SU(2) soliton into SU(3)
\begin{equation}
  \label{eq:embed}
U_{\mathrm{SU(3)}} = \left(\begin{array}{lr} U_{\mathrm{SU(2)}} & 0
    \\ 0 & 1   \end{array}\right)
\end{equation}
with the SU(2) pion field $\pi(x)$ as
\begin{equation}
U^{\gamma_5} = \exp(i\gamma^5 \tau^i\pi^i(x)) =
\frac{1+\gamma^5}{2}U + \frac{1-\gamma^5}{2} U^\dagger.
\end{equation}
The integration over the pion field in Eq.(\ref{eq:part}) can be done
by using the saddle-point approximation in the large $N_c$ limit due
to the $N_c$ factor in Eq.(\ref{eq:echl}). In order to find the pion
field that minimizes the action in Eq.(\ref{eq:echl}) and to calculate 
Eq.(\ref{matelem}), we make the following Ans\"atze.  The SU(2)
pion field $U$ has the most symmetric form known as the hedgehog form:
\begin{equation}
  \label{eq:hedgehog}
U_{\mathrm{SU2}}=\exp[i\gamma_5 \hat{\bm n}\cdot\bm\tau P(r)],
\end{equation}
where $P(r)$ is the radial pion profile function.

The baryon state is defined as an Ioffe-type current consisting of
$N_c$ valence quarks~\cite{Christov:1995vm}:
\begin{eqnarray}
|B(p)\rangle & = &
\lim_{x_{4}\to-\infty}\,\frac{1}{\sqrt{\mathcal{Z}}}\,
e^{ip_{4}x_{4}}\,\int d^{3}{\bm x}\, e^{i\,{\bm p}\cdot{\bm x}}\,
J_{B}^{\dagger}(x)\,|0\rangle,\cr
J_{B}(x) & = &
\frac{1}{N_{c}!}\,\Gamma_{B}^{b_{1}\cdots b_{N_{c}}}\,
\varepsilon^{\beta_{1}\cdots\beta_{N_{c}}}\,\psi_{\beta_{1}b_{1}}(x)
\cdots\psi_{\beta_{N_{c}}b_{N_{c}}}(x),
\end{eqnarray}
where the matrix $\Gamma_{B}^{b_{1}...b_{N_{c}}}$ carries the
hyper-charge $Y$, isospin $I,I_{3}$ and spin $J,J_{3}$ quantum numbers
of the baryon. The $b_{i}$ and $\beta_{i}$ denote the spin-flavor- and 
color-indices, respectively. Thus, we can write the baryonic matrix
element of the quark current $\mathcal{J}^{\mu\chi}$ as follows:
\begin{eqnarray}
 \langle B_{2}(p_{2})|\mathcal{J}^{\mu \chi}(0)|B_{1}(p_{1}\rangle
& = &  \frac{1}{\mathcal{Z}}\lim_{T\to\infty}
e^{-ip_{2}^{4}\frac{T}{2} + ip_{1}^{4}\frac{T}{2}} \int  d^{3}{\bm
  x}^{\prime} d^{3}{\bm x}e^{i{\bm p}_{1}\cdot{\bm x}-i{\bm
     p}_{2}\cdot{\bm x}^{\prime}}\cr
&\times& \int\mathcal{D}U\mathcal{D}\psi^{\dagger}\mathcal{D}\psi
 J_{B^{\prime}} \left(\frac{T}{2},{\bm
     x}^{\prime}\right)\mathcal{J}^{\mu \chi}(0)
J_{B}^{\dagger} \left(-\frac{T}{2},{\bm x}\right)\cr
&&\times \exp{\left[-\int d^{4}x\,\psi^\dagger i D(U) \psi\right]}.
\label{eq: path-integral matrix-element}
\end{eqnarray}
Taking $\mathcal{J}^{\mu \chi}(0)=1$ in
Eq.(\ref{eq: path-integral matrix-element}), we get the
nucleon correlation function from which we can obtain
the expression for the classical nucleon mass in the limit of
$T\to\infty$.  The self-consistent soliton profile $P_c(r)$ in the
saddle-point approximation is found by minimizing the nucleon mass in
the $\chi$QSM, i.e., by solving numerically the equation of motion
coming from $\delta S_{eff}/\delta P(r)=0$.

Since the soliton does not have good quantum numbers for rotations as
well as translations, we need to quantize it. This can by done by the 
zero-mode quantization as follows:
\begin{equation}
U(\bm x, t) = A(t)U_c(\bm x - \bm z(t))A^\dagger (t),
\end{equation}
where $A(t)$ denotes a unitary time-dependent SU(3) collective
orientation matrix and $\bm z(t)$ stands for the time-dependent
translation of the center of mass of the soliton in coordinate
space. A detailed formalism can be found in
Refs.\cite{Christov:1995vm,Kim:eleff}. Taking the zero-modes into
account, we find the Dirac operator in Eq.(\ref{eq:Dirac}) in the
following form:
\begin{equation}
D(U)=T_{z(t)}A(t) \left[ D(U_{c}) + i\Omega(t) -
  \dot{T}_{z(t)}^{\dagger}T_{z(t)}-i\gamma^{4}A^{\dagger}(t)\delta m
  A(t) \right]T_{z(t)}^{\dagger}A^{\dagger}(t),
\end{equation}
where the $T_{z(t)}$ denotes the translational unitary operator and
the $\Omega(t)$ represents the soliton angular velocity
defined as
\begin{equation}
\Omega=-iA^{\dagger}\dot{A}=-\frac{i}{2}\textrm{Tr} (
A^{\dagger}\dot{A}\lambda^{\alpha})\lambda^{\alpha}=\frac{1}{2}
\Omega_{\alpha}\lambda^{\alpha}.
\end{equation}
Assuming that the soliton rotates and moves slowly, we can treat the
$\Omega(t)$ and $\dot{T}_{z(t)}^{\dagger}T_{z(t)}$ perturbatively.
Moreover, the strange current quark mass $\hat{m}$ is regarded as a
small parameter, so that we also deal with $\delta m$
perturbatively. The collective baryon wavefunction on the level of
Eq.(\ref{eq: path-integral matrix-element}) is then introduced as
\begin{equation}
\psi_{(\mathcal{R}^{*};Y^{\prime}JJ_{3})}^{(\mathcal{R};YII_{3})}(A)
:= \lim_{T\mapsto\infty}\frac{1}{\sqrt{\mathcal{Z}}}e^{-p^{4\prime}T/2}\int
d^{3}{\bm u}^{\prime}e^{i{\bm p}^{\prime}\cdot{\bm u}^{\prime}}
(\Gamma_{B}^{b_{1}...b_{N_{c}}})^*\Pi_{l=1}^{N_{c}}\big[
\varphi^{\dagger}_{v,b_{l}}({\bm u}^{\prime})A^\dagger \big].
\end{equation}
Having expanded and quantized the soliton, we obtain the following
collective Hamiltonian~\cite{CQSM_Quantization}:
\begin{equation}
  \label{eq:Ham}
H_{\textrm{coll}}=H_{\mathrm{sym}} + H_{\mathrm{sb}} \,\,\, ,
\end{equation}
where $H_{\mathrm{sym}}$ and $H_{\mathrm{sb}}$ represent the SU(3)
symmetric and symmetry-breaking parts, respectively:
\begin{eqnarray}
H_{\mathrm{sym}} &=& M_{c} + \frac{1}{2I_{1}}\sum_{i=1}^3 J_{i}J_{i} +
\frac{1}{2I_{2}} \sum_{a=4}^7 J_{a} J_{a} ,\cr
H_{\mathrm{sb}} &=& \frac{1}{\overline{m}} M_{1}
\Sigma_{SU(2)} + \alpha D_{88}^{(8)}(A) + \beta Y +
\frac{\gamma}{\sqrt{3}} D_{8i}^{(8)}(A)J_{i}.
\end{eqnarray}
The $M_c$ denotes the mass of the classical soliton and $I_{i}$ and
$K_{i}$ are the moments of inertia of the
soliton~\cite{Christov:1995vm}, of which the corresponding
expressions can be found in Ref.~\cite{Blotz_mass_splittings}
explicitly.  The components $J_i$  denote the spin generators and
$J_a$ correspond to those of right rotations in flavor SU(3) space.  The
$\Sigma_{\mathrm{SU(2)}}$ is the SU(2) pion-nucleon sigma term. The
$D_{88}^{(8)}(A)$ and $D_{8i}^{(8)}(A)$ stand for the SU(3) Wigner $D$
functions in the octet representation.  The $Y$ is the hypercharge
operator.  The parameters $\alpha$, $\beta$, and $\gamma$ in the
symmetry-breaking Hamiltonian are expressed, respectively, as follows:
\begin{equation}
\alpha = \frac{1}{\overline{m}} \frac{1}{\sqrt{3}} M_{8}
\Sigma_{SU(2)} - \frac{N_{c}}{\sqrt{3}} M_{8} \frac{K_{2}}{I_{2}},
\;\;\;\;
\beta = M_{8} \frac{K_{2}}{I_{2}}\sqrt{3},\;\;\;\;
\gamma = -2\sqrt{3}M_{8}\left(\frac{K_{1}}{I_{1}} -
  \frac{K_{2}}{I_{2}}\right).
\end{equation}
The collective wavefunctions of the Hamiltonian in Eq.(\ref{eq:Ham})
can be found as the SU(3) Wigner $D$ functions in representation
$\mathcal{R}$:
\begin{equation}
  \label{eq:Wigner}
\langle A|\mathcal{R},B(YII_{3},Y^{\prime}JJ_{3}) \rangle =
\Psi_{(\mathcal{R}^{*};Y^{\prime}JJ_{3})}^{(\mathcal{R};YII_{3})}(A) =
\sqrt{\textrm{dim}(\mathcal{R})}\,(-)^{J_{3}+Y^{\prime}/2}\,
D_{(Y,I,I_{3})(-Y^{\prime},J,-J_{3})}^{(\mathcal{R})*}(A).
\end{equation}
The $Y'$ is related to the eighth component of the angular velocity
$\Omega$ that is due to the presence of the discrete valence quark
level in the Dirac-sea spectrum.  Its presence has no effect on the
chiral field, so that it is constrained to be $Y'=-N_c/3=-1$.  In fact,
this constraint allows us to have only the SU(3) representations with
zero triality.

Due to flavor SU(3) symmetry breaking the collective baryon states are not in a pure
representation but get mixed with other representations.
This can be treated by first-order perturbation for
the collective Hamiltonian:
\begin{equation}
|B_{\mathcal{R}}\rangle = |B_{\mathcal{R}}^{\mathrm{sym}} \rangle -
\sum_{\mathcal{R}^{\prime}\neq\mathcal{R}} |B_{\mathcal{R}^{\prime}}
\rangle \frac{\langle B_{\mathcal{R}^{\prime}}|\,
  H_{\textrm{sb}}   \,|B_{\mathcal{R}}\rangle}{M(\mathcal{R}^{\prime}) -
  M(\mathcal{R})}\,.
\label{wfc}
\end{equation}
Solving Eq.(\ref{wfc}), we obtain the collective wavefunctions for
the baryon octet:
\begin{equation}
|B_{8}\rangle  =  |8_{1/2},B\rangle + c_{\overline{10}}^{B}
|\overline{10}_{1/2}, B\rangle + c_{27}^{B}|27_{1/2},B\rangle,
\label{B8}
\end{equation}
where the mixing coefficients $c_{\overline{10}}^{B}$ and $c_{27}^{B}$
are defined as
\begin{equation}
c_{\overline{10}}^{B}=c_{10}\,\Big[\sqrt{5},0,\sqrt{5},0\Big],
\,\,\,\,\,\,\,\,\,\,
c_{27}^{B}=c_{27}\,\Big[\sqrt{6},3,2,\sqrt{6}\Big]
\end{equation}
in the basis $[N,\Lambda,\Sigma,\Xi]$ with
\begin{equation}
c_{\overline{10}}=-\frac{I_{2}}{15}\Big(\alpha+\frac{1}{2}\gamma\Big),
 \,\,\,\,\,\,\,\,\,\,
c_{27}=-\frac{I_{2}}{25}\Big(\alpha-\frac{1}{6}\gamma\Big).
\label{mixing_coeff}
\end{equation}
We are now in a position to evaluate the baryonic matrix elements such
as Eq.(\ref{matelem}), i.e. to solve
Eq.(\ref{eq: path-integral matrix-element}) with a certain expression
of $\mathcal{J}^{\mu \chi}(0)$ given.  We have now the general
expression for the baryonic matrix elements as follows:
\begin{equation}
\langle B_{2}(p_{2})| \psi^{\dagger}(0) \mathcal{O}^{\mu\chi} \psi(0)|
B_{1}(p_{1})\rangle = \int dA\int d^{3}z\,\,
e^{i{\bm q}\cdot{\bm z}}\Psi_{B_{2}}^{*}(A)\mathcal{G}^{\mu\chi}({\bm
  z})
\Psi_{B_{1}}(A)e^{S_{eff}},
\label{general model eq}
\end{equation}
where $dA$ and $d^3z$ arise from the zero-mode quantizations and the
expression $\mathcal{G}^{\mu\chi}({\bm z})$ contains the specific form
factor parts. We have used again the saddle-point approximation and
expanded the Dirac operator with respect to $\Omega$ and $\delta m$ to
the linear order and $ \dot{T}_{z(t)}^{\dagger}T_{z(t)}$ to the zeroth
order.  Defining $\mathcal{J}^{\mu \chi}(0)$ and contracting the
Lorentz index in Eq.(\ref{general model eq}) according to
Eqs.(\ref{GE},\ref{GM},\ref{GA}), we get the final expressions in the
$\chi$QSM for the vector form factors as follows:
\begin{eqnarray}
G_{E}^{\chi}(Q^{2}) & = & \int d^{3}z\, j_{0}(|{\bm q}||{\bm
  z}|)\,\langle B_{2}|\mathcal{G}_{E}^{\chi}({\bm
  z})|B_{1}\rangle,
\label{eq:G_E cqsm}\\
G_{M}^{\chi}(Q^{2}) & = & M_{p}\int
d^{3}z\frac{j_{1}(|{\bm q}||{\bm z}|)}{|{\bm q}||{\bm z}|} \, \langle  
B_{2}|\mathcal{G}_{M}^{\chi}({\bm z})|B_{1}\rangle,
\label{eq:G_M cqsm}
\end{eqnarray}
and axial-vector form factors as
\begin{equation}
g_{1}^{\chi}(Q^{2})=\int
d^{3}z\,\Big[j_{0}(|{\bm z}||{\bm q}|)\,\langle
B_{2}|\mathcal{G}_{10}^{\chi}({\bm z})|B_{1}\rangle -
\sqrt{2\pi}j_{2}(|{\bm z}||{\bm q}|)\,\langle
B_{2}|\{Y_{2}\otimes\mathcal{G}_{1}^{\chi}({\bm z})\}_{10} |
B_{1}\rangle\Big],
\label{eq:G_A cqsm}
\end{equation}
where $j_{0,1,2}$ denote the spherical Bessel functions.  The
expressions $\mathcal{G}_{E}^{\chi}({\bm z})$,
$\mathcal{G}_{M}^{\chi}({\bm z})$ and $\mathcal{G}_{10}^{\chi}({\bm
  z}),\{Y_{2}\otimes\mathcal{G}_{1}^{\chi}({\bm z})\}_{10}$
can be found in Appendix~\ref{app:a}.  The expansion in $\Omega$ and
$\delta m$ provides the following structure of the form factor
structure:
\begin{eqnarray}
G^{B_1 B_2}_{E,M,A}(Q^{2}) & = &
G_{E,M,A}^{(\Omega^{0},m_{s}^{0})}(Q^{2})+G_{E,M,A}^{(\Omega^{1},m_{s}^{0})}(Q^{2})
+G_{E,M,A}^{(m_{s}^{1}),\textrm{op}}(Q^{2})+G_{E,M,A}^{(m_{s}^{1}),\textrm{wf}}(Q^{2}),
\label{eq:cqsm ff}
\end{eqnarray}
where the first term corresponds to the leading order ($\Omega^{0},m_{s}^{0}$),
the second one to the first $1/N_{c}$ rotational corrections
($\Omega^{1},m_{s}^{0}$), the third to the linear $m_{s}$ corrections
coming from the operator, and the last one to the wavefunction
corrections, respectively.

In the whole approach we take only orders of $\mathcal{O}(m_{s}^{0})$
and $\mathcal{O}(m_{s}^{1})$ into account for the expression
$\mathcal{G}^{\mu\chi}$ and consider only the zeroth order
$\mathcal{O}(m_{s}^{0})$ for the expression $e^{i{\bm q}\cdot{\bm z}}$
in Eq.(\ref{general model eq}).  In the $\chi$QSM, we may write the
masses $M_{B_2}$ of the baryon octet as
\begin{equation}
M_{B_2}=M_{B_1}+\textrm{const}\cdot m_{s}
\label{mass in CQSM}
\end{equation}
with $M_{B_1}\sim\mathcal{O}(N_{c})$, so that the second term is of
order $\mathcal{O}(N_{c}^{0},m_{s}^{1})$.  Inserting
Eq.(\ref{mass in CQSM}) in the definition of the momentum transfer
in Eq.(\ref{vecqdef}) and neglecting all terms of higher order than
$\mathcal{O}(m_{s}^{0})$, we get
\begin{equation}
{\bm q}^{2}\stackrel{ N_{c}\to\infty}{=}Q^{2},
\label{vecq in large NC}
\end{equation}
which means that for each transition form factor the momentum
dependence turns out to be
\begin{equation}
j_{0,1,2}(|{\bm q}||{\bm z}|)\to
j_{0,1,2}(\sqrt{Q^{2}}|{\bm z}|)\,\,\,.
\end{equation}
In order to get the baryonic matrix element in the form of
Eq.(\ref{general model eq}), we have used the large $N_c$
limit.  Taking also the limit of $N_{c}\to\infty$ on the RHS of  
Eqs.(\ref{eq:eff Ge},\ref{eq:eff Gm}), we obtain the following
relations:
\begin{eqnarray}
E_{B} & = & M_{B}+\frac{{\bm p}}{2M_{B}}+\mathcal{O}(N_{c}^{-2}),\cr
\sqrt{\frac{E_{B}+M_{B}}{2M_{B}}} & = & 1+\mathcal{O}(N_{c}^{-2}),\cr
\frac{1}{M_{B_1}}\frac{N{\bm q}^{2}}{E_{2}+M_{B_2}} & = &
\mathcal{O}(N_{c}^{-2})+\mathcal{O}(N_{c}^{-1},m_{s}^{1}),\cr
\frac{q^{0}}{M_{B_1}}N & = &
\mathcal{O}(N_{c}^{-2})+\mathcal{O}(N_{c}^{-1},m_{s}^{1}), \cr
\frac{2M_{p}N}{E_{2}+M_{B_2}} & = & \frac{2M_{p}(1 +
  \mathcal{O}(N_{c}^{-2}))}{ 2M_{B_2}(1 + \mathcal{O}(N_{c}^{-2}))} =
\frac{M_{p}}{M_{B_2}},
\end{eqnarray}
so that the vector form factors calculated in the $\chi$QSM via
Eqs.(\ref{eq:G_E cqsm},\ref{eq:G_M cqsm}) corresponds to
\begin{eqnarray}
G_{E}^{B_{1}B_{2}}(Q^{2}) & = & f_{1}^{B_{1}B_{2}}(Q^{2})
\label{eq:Ge after Nc}\\
G_{M}^{B_{1}B_{2}}(Q^{2}) & = & \frac{M_{p}}{M_{B_2}}
\Big[f_{1}^{B_{1}B_{2}}(Q^{2}) + f_{2}^{B_{1}B_{2}}(Q^{2})
\frac{M_{B_1}+M_{B_2}}{M_{B_1}} \Big].
\label{eq:Gm  after Nc}
\end{eqnarray}
The factor of $1/N$ in Eq.(\ref{GA}) reduces to $1$ because of
the same arguments.

In the $\chi$QSM Hamiltonian of Eq.(\ref{eq:diracham}), the
constituent quark mass $M$ is the only free parameter and
$M=420\,\textrm{MeV}$ is known to reproduce very well experimental
data~\cite{SilvaKUG:2005,Christov:eleff,Silva:2005qm,Kim:eleff}.
Though the $M=420$ MeV yields the best results for the baryon octet,
we will present also those for $M=400\,\textrm{MeV}$ and
$M=450\,\textrm{MeV}$ to see the $M$ dependence of the results in this
work.  Throughout this work the strange current quark mass is fixed to
$m_{\mathrm{s}}=180\,\textrm{MeV}$.  In order to tame the divergent
quark loops, we employ in this work the proper-time regularization.
The cut-off parameter and $\overline{m}$ are fixed for a given $M$
to the pion decay constant $f_{\pi}$ and $m_\pi$, respectively.  The
numerical results for the moments of inertia and mixing coefficients
are summarized in Table~\ref{Numerical-values} for $M=420$ MeV.
\begin{table}[t]
\caption{\label{Numerical-values} Moments of inertia and mixing
coefficients for $M=420\,\textrm{MeV}$.}
\begin{center}
\begin{tabular}{ccccccc}
\hline
$I_{1}\,[\textrm{fm}]$&
$I_{2}\,[\textrm{fm}]$&
$K_{1}\,[\textrm{fm}]$&
$K_{2}\,[\textrm{fm}]$&
$\Sigma_{\pi N}\,[\textrm{MeV}]$ &
$c_{\overline{10}}$&
$c_{27}$
\\ \hline
$1.06$&
$0.48$&
$0.42$&
$0.26$&
$41$&
$0.037$&
$0.019$
\tabularnewline
\hline
\end{tabular}
\end{center}
\end{table}
The results in Table~\ref{Numerical-values} are obtained with
the same parameters used in previous
works~\cite{Silva:2005qm,silva_data:1999,silva_data:2002}.  We want to
emphasize that all model parameters are the same as before.  In
previous works, the axial-vector form factors for the
nucleon were already calculated.  The axial-vector constants
$g_{A}^{3},g_{A}^{8}$ were found to be $g_{A}^{3}=1.176$ and
$g_{A}^{8}=0.36$ which is in very good agreement with experimental
data $g_{A}^{3}=1.267\pm0.0029$~\cite{PDG:2006} and
$g_{A}^{8}=0.338\pm0.15$~\cite{exp_axial_data_g8}. With these
numerical parameters, the $\chi$QSM yields the masses of the octet
baryons in unit of MeV as Ref.~\cite{decay_Theta}:
\begin{equation}
M_N=1001 (939), \,\,\,\, M_\Lambda=1124 (1116), \,\,\,\, M_\Sigma=1179
(1189), \,\,\,\, M_\Xi= 1275 (1318),
\end{equation}
where the numbers in the parentheses are the experimental values of
the Particle Data Group~\cite{PDG:2006}.  The $\chi$QSM values were
obtained by first calculating the hyper-charge splittings with
Eq.(\ref{eq:Ham}) and using the experimental octet mass center of
$M_8=(M_\Lambda+M_\Sigma)/2=1151.5$MeV.
\section{Results and Discussion}
In the present work, we consider linear $m_{s}$ corrections.
According to the Ademollo-Gatto theorem~\cite{AG_theorem}
SU(3) symmetry-breaking effects contribute to the matrix elements of
the vector current between the states in the same SU(3)
multiplet earliest at second order in $m_{s}$.
Since we restrict ourselves in the present investigation to first
order in $m_{s}$, we have for $f_{1}^{B_{1}B_{2}}(0)$ only the SU(3)
symmetric results.  The SHD constants in the $\chi$QSM were already
investigated in Ref.~\cite{CQSM_semilep1998}.  The constants of
$g_{1}(0)/f_{1}(0)$ were also discussed in the $\chi$QSM formulated
in a Fock representation on the light cone \cite{lorce_semilep}.  In
the present work we will extend the earlier
work~\cite{CQSM_semilep1998} by considering the full form factors of
$f_{1}(Q^2)$,  $f_{2}(Q^2)$ and $g_{1}(Q^2)$ up to $Q^2\leq 1
\textrm{GeV}^2$ and also taking into account the symmetry-conserving 
quantization. The symmetry-conserving quantization was found after the
publication of Ref.~\cite{CQSM_semilep1998} and does also have effects
on the SHD constants.
\subsection{SHD vector form factors}
We now present the results for the vector transition form factors
$f^{B_1B_2}_1(Q^2)$ and $f^{B_1B_2}_2(Q^2)$ in the $\chi$QSM, for which
we have the following relations from Eq.(\ref{eq:Ge after Nc}):
\begin{equation}
f_{1}^{B_{1}B_{2}}(Q^2)=G_{E}^{B_{1}B_{2}}(Q^2),
\,\,\,\,\,\,\,\,\,\,\,\,\,
f_{2}^{B_{1}B_{2}}(Q^2)=\frac{M_{B_1}}{M_{B_1}+M_{B_2}}\Big[G_{M}^{B_{1}B_{2}}(Q^2)
\frac{M_{B_2}}{M_{p}}-f_{1}^{B_{1}B_{2}}(Q^2)\Big].
\label{f1f2}
\end{equation}
We first discuss the results of the SHD constants at $Q^{2}=0$.  Due
to the expressions
\begin{equation}
\int d^{3}z\,\mathcal{B}({\bm z})=3,\;\;\;\frac{1}{I_{i}}\int
d^{3}z\,\mathcal{I}_{i}({\bm z})=1,\;\;\;\frac{1}{K_{i}}\int
d^{3}z\,\mathcal{K}_{i}({\bm z})=1,\;\;\;\int
d^{3}z\,\mathcal{C}({\bm z})=0,
\end{equation}
we see that Eq.(\ref{eq:G_E cqsm}) with Eq.(\ref{eq:Ele dens}) and the
corresponding coefficients in Appendix~\ref{CGs} give the following
values:
\begin{eqnarray}
f_{1}^{np}(0)=1, & \,\,\,\,\,\,\,\,\,\,\,\,\,\,\,\,\,\, &
f_{1}^{\Sigma^{-}n}(0)=1,\cr
f_{1}^{\Sigma^{-}\Lambda}(0)=0, &  & f_{1}^{\Lambda
  p}(0)=\sqrt{\frac{3}{2}},\cr
f_{1}^{\Sigma^{-}\Sigma^{0}}(0)=\sqrt{2}, &  &
f_{1}^{\Xi^{-}\Sigma^{0}}(0)=\frac{1}{\sqrt{2}},\cr
f_{1}^{\Xi^{-}\Xi^{0}}(0)=1, &  &
f_{1}^{\Xi^{-}\Lambda}(0)=\sqrt{\frac{3}{2}}.
\end{eqnarray}
Since we only take linear $m_s$ corrections into
account the $m_s$ corrections both from the operator and from the
wavefunction add up to zero at $Q^2=0$ which is the consequence of the
Ademollo-Gatto theorem.

In the case of the SHD constants $f_2(0)$, Eq.(\ref{eq:G_M cqsm}) for
$G_{M}^{B_{1}B_{2}}(0)$ is involved. This expression contains
explicitly a factor of $M_p$.  The proton mass in the $\chi$QSM turns
out to be $1.36$ times larger than the experimental proton
mass, i.e. $M_{p}^{\chi\mathrm{QSM}} = 1.36M_{p}^{\mathrm{exp}}$. This is a well 
understood property of the $\chi\mathrm{QSM}$ \cite{Pobylitsa:1992bk},
whose consequences we cure as usual in the following way. 
We will calculate Eq.(\ref{eq:G_M cqsm}), which is a pure model
expression, with the soliton-nucleon mass.  However, having determined
the $G_{M}^{B_{1}B_{2}}(0)$, we will then use experimental values for
further calculations, such as calculating $f_{2}^{B_{1}B_{2}}(0)$ in
Eq.(\ref{f1f2}).  The used experimental masses are given in units of
GeV as follows:
\begin{eqnarray}
M_{p}=M_N=0.939, & \,\,\,\,\,\,\,\,\,\,\,\,\,\: &
M_{\Lambda}=1.116,\cr
M_{\Sigma}=1.186, &  & M_{\Xi}=1.318.
\label{Bar masses}
\end{eqnarray}
Results for the constants $f_{1}^{B_{1}B_{2}}(0)$, $G_{M}^{B_{1}B_{2}}(0)$
and $f_{2}^{B_{1}B_{2}}(0)$ are given in Table~\ref{tab f1gmf2} for
the constituent quark mass of $M=420$MeV. We find that the constants
are not sensitive to $M$.  In Ref.\cite{CQSM_semilep1998} it was found
that linear $m_s$ corrections to the form factor $f_2$ are sizable for
some transitions, explicitly for the transitions $\Sigma^- \to
\Sigma^0$ and $\Xi^- \to \Sigma^0$.  Due to the application of the
symmetry-conserving quantization the absolute values of the transition
constants are changing but the effects of the $m_s$ corrections on this
work are comparable to those in Ref.~\cite{CQSM_semilep1998}.
In the case of $G_{M}(0)$, we have used the soliton-nucleon mass in
Eq.(\ref{eq:Gm  after Nc}), while we have employed experimental masses
for Eq.(\ref{f1f2}). 
\begin{table}[ht]
\caption{\label{tab f1gmf2}
Results for $f_{1}^{B_{1}B_{2}}(0)$, $G_{M}^{B_{1}B_{2}}(0)$ and
$f_{2}^{B_{1}B_{2}}(0)$ in the self-consistent $\chi$QSM for the
constituent quark mass of $420$ MeV.  We show the results with and
without linear $m_s$ corrections. The SHD constants
$f_{1}^{B_{1}B_{2}}(0)$ do not acquire any linear $m_s$
corrections. The superscript $[\mathrm{SU(3)}]$ indicates the SU(3) 
symmetry conserving approximation. The final results are given in the
column $f_{1}(0)$ and in the columns with superscript
$[m_{s}^{0}+m_{s}^{1}]$. The $SU(3)$ anomalous magnetic moments of
the nucleons in the $\chi$QSM are obtained as $\kappa_p=1.47$ and
$\kappa_n=-1.64$.}
\begin{tabular}{c|cc|ccc}
\hline
&
$G_{M}^{\mathrm{SU(3)}}(0)$&
$G_{M}^{m_{s}^{0}+m_{s}^{1}}(0)$&
$f_{1}(0)$&
$f_{2}^{\mathrm{SU(3)}}(0)$&
$f_{2}^{m_{s}^{0}+m_{s}^{1}}(0)$\tabularnewline
\hline
$n\to p$&
$4.10$&
$4.13$&
$1$&
$\frac{1}{2}(\kappa_p -\kappa_n)=1.55$&
$1.57$\tabularnewline
$\Sigma^{-}\to\Lambda$&
$2.00$&
$2.02$&
$0$&
$-\sqrt{\frac{3}{8}}\kappa_n=1.00$&
$1.24$\tabularnewline
$\Sigma^{-}\to\Sigma^{0}$&
$2.33$&
$2.35$&
$\sqrt{2}$&
$\frac{1}{\sqrt{2}}(\kappa_p +\frac{1}{2}\kappa_n)=0.46$&
$0.78$\tabularnewline
$\Xi^{-}\to\Xi^{0}$&
$-0.80$&
$-0.81$&
$1$&
$\frac{1}{2}(\kappa_p + 2 \kappa_n)=-0.90$&
$-1.07$\tabularnewline
\hline
$\Sigma^{-}\to n$&
$-0.80$&
$-0.73$&
$1$&
$\frac{1}{2}(\kappa_p + 2 \kappa_n)=-0.90$&
$-0.96$\tabularnewline
$\Lambda\to p$&
$3.02$&
$2.83$&
$\sqrt{\frac{3}{2}}$&
$\sqrt{\frac{3}{8}}\kappa_p=0.90$&
$0.87$\tabularnewline
$\Xi^{-}\to\Sigma^{0}$&
$2.90$&
$2.72$&
$\frac{1}{\sqrt{2}}$&
$\frac{1}{2\sqrt{2}}(\kappa_p - \kappa_n)=1.10$&
$1.44$\tabularnewline
$\Xi^{-}\to\Lambda$&
$1.02$&
$0.97$&
$\sqrt{\frac{3}{2}}$&
$\sqrt{\frac{3}{8}}(\kappa_p +\kappa_n)=-0.10$&
$-0.04$\tabularnewline
\hline
\end{tabular}
\end{table}

We compare our present results of the ratios
$(f_{2}(0)/f_{1}(0))^{B_{1}B_{2}}$ with those of
Refs.\cite{Cabibbo2003, CQSM_semilep1998}. In Ref.~\cite{Cabibbo2003}, 
the following relations are used:
\begin{eqnarray*}
\Big(\frac{f_{2}(0)}{f_{1}(0)}\Big)^{np}=\frac{M_{n}}{M_{p}}
\frac{(\kappa_p-\kappa_n)}{2},
& \,\,\,\,\,\,\,\,\, &
\Big(\frac{f_{2}(0)}{f_{1}(0)}\Big)^{\Sigma^{-}n} =
\frac{M_{\Sigma^{-}}}{M_{p}}\frac{(\kappa_p+2\kappa_n)}{2},\cr
\Big(\frac{f_{2}(0)}{f_{1}(0)}\Big)^{\Sigma^{-}\Lambda} =
-\frac{M_{\Sigma^{-}}}{M_{p}}\frac{\kappa_n}{2}\sqrt{\frac{3}{2}}, &
& \Big(\frac{f_{2}(0)}{f_{1}(0)}\Big)^{\Lambda
  p}=\frac{M_{\Lambda}}{M_{p}}\frac{\kappa_p}{2},\cr
\Big(\frac{f_{2}(0)}{f_{1}(0)}\Big)^{\Sigma^{-}\Sigma^{0}} =
\frac{M_{\Sigma^{-}}}{M_{p}}\frac{(2\kappa_p+\kappa_n)}{4}, &  &
\Big(\frac{f_{2}(0)}{f_{1}(0)}\Big)^{\Xi^{-}\Sigma^{0}} =
\frac{M_{\Xi^{-}}}{M_{p}}\frac{(\kappa_p-\kappa_n)}{2},\cr
\Big(\frac{f_{2}(0)}{f_{1}(0)}\Big)^{\Xi^{-}\Xi^{0}} =
\frac{M_{\Xi^{-}}}{M_{p}}\frac{(\kappa_p+2\kappa_n)}{2},
&  &
\Big(\frac{f_{2}(0)}{f_{1}(0)}\Big)^{\Xi^{-}\Lambda} =
-\frac{M_{\Xi^{-}}}{M_{p}}\frac{(\kappa_p+\kappa_n)}{2},
\end{eqnarray*}
where the experimental anomalous magnetic moments are given as 
$\kappa_p=1.793\,\mu_N$ and $\kappa_n=-1.91\,\mu_N$ with $\mu_N$ being
the nuclear magneton.  In Table~\ref{resf2/f1}, we
compare our final results for $(f_{2}/f_{1})^{B_{1}B_{2}}$ with those
of the Cabibbo model in $SU(3)$ as well as with those in
Ref.\cite{CQSM_semilep1998}.  As for the constants $f_2(0)$, the
ratios $(f_{2}/f_{1})^{B_{1}B_{2}}$ show strong $m_s$ contributions as
found in Ref.~\cite{CQSM_semilep1998}. Prominent are 
the transitions $\Sigma^- \to \Sigma^0$ and $\Xi^- \to \Sigma^0$.
\begin{table}[ht]
\caption{\label{resf2/f1}
Results for the ratios $(f_{2}/f_{1})^{B_{1}B_{2}}$ in the
self-consistent $\chi$QSM for the constituent quark mass $420$
MeV. In the second column, the results of the Cabibbo
model~\cite{Cabibbo2003} (CM) are listed, while in the third column
those of the $\chi$QSM~\cite{CQSM_semilep1998} without the
symmetry-conserving quantization are presented.  The next two columns
list the results of the present work without and with $m_s$
corrections. The final results are given by $\chi
\textrm{QSM}_{m_{s}^{0}+m_{s}^{1}}$. Experimental data are taken from 
Refs.~\cite{Sigm2n1988,Cabibbo2003,
Lam2p1990,BGHORS:1983,KTeV2001}. \protect \\}
\begin{tabular}{c|cc|cc|c}
\hline
$f_{2}(0)/f_{1}(0)$&
CM~\cite{CQSM_semilep1998} &
$m_{s}^{0}+m_{s}^{1}$~\cite{CQSM_semilep1998} &
$\chi \textrm{QSM}_{SU(3)}$&
$\chi \textrm{QSM}_{m_{s}^{0}+m_{s}^{1}}$&
Experiment\tabularnewline
\hline
$n\to p$&
$1.85$&
$1.85$&
$1.55$&
$1.57$&
\tabularnewline
$\Sigma^{-}\to\Lambda$
\footnote{Since $f_{1}^{\Sigma^{-}\Lambda}(0)=0$, we list $f_{2}(0)$
instead $f_{2}/f_{1}$.}&
$1.17$&
$1.33$&
$1.00$&
$1.24$&
\tabularnewline
$\Sigma^{-}\to\Sigma^{0}$&
$0.42$&
$0.52$&
$0.33$&
$0.55$&
\tabularnewline
$\Xi^{-}\to\Xi^{0}$&
$-1.01$&
$\cdots$&
$-0.90$&
$-1.08$&
\tabularnewline
\hline
$\Sigma^{-}\to n$&
$-1.02$&
$-1.01$&
$-0.90$&
$-0.96$&
$-0.96\pm0.15$\tabularnewline
$\Lambda\to p$&
$0.90$&
$0.79$&
$0.73$&
$0.71$&
$0.15\pm0.30$; $\frac{1.34\pm0.20}{1.238\pm0.024}$
\tabularnewline
&&&&& $1.32\pm0.81$ \tabularnewline
$\Xi^{-}\to\Sigma^{0}$&
$1.85$&
$1.73$&
$1.56$&
$2.02$&
$2.0^{\pm 1.2(\textrm{stat})}_{\pm0.5(\textrm{syst})}$~\footnote{The data
correspond to the $\Xi^0 \to \Sigma^+$ transition.}
\tabularnewline
$\Xi^{-}\to\Lambda$&
$-0.06$&
$-0.09$&
$-0.08$&
$-0.02$&
\tabularnewline
\hline
\end{tabular}
\end{table}
Experimental data on the SHD $(f_{2}(0)/f_{1}(0))$ ratios are
available only for three transitions: $\Sigma^{-}\to n$,
$\Lambda\to p$ and $\Xi^{0}\to\Sigma^{+}$~\cite{Sigm2n1988,
Cabibbo2003,Lam2p1990,BGHORS:1983,KTeV2001}.  We can use the following
isospin relations given in Appendix~\ref{CGs}:
\begin{equation}
\Xi^{0}\to\Sigma^{+}=\sqrt{2}\Big[\Xi^{-}\to\Sigma^{0}\Big],
\end{equation}
for both $f_{2}^{\Xi\Sigma}(0)$ and $f_{1}^{\Xi\Sigma}(0)$, so that we
have the results as listed in Table~\ref{resf2/f1}:
\begin{equation}
f_{1}^{\Xi^{0}\Sigma^{+}}(0)=1,\;\;\;
f_{2\,\,\chi\mathrm{QSM}}^{\Xi^{0}\Sigma^{+}\,\,\mathrm{SU(3)}}(0) =
1.56,\;\;\;
f_{2\,\,\chi\mathrm{QSM}}^{\Xi^{0}\Sigma^{+}\,\,
  m_{s}^{0}+m_{s}^{1}}(0) = 2.02.
\end{equation}
Comparing our results with the experimental data as shown in
Table~\ref{resf2/f1}, we find that they are in good agreement with
the data.

\begin{figure}
\begin{center}
\includegraphics[scale=0.5]{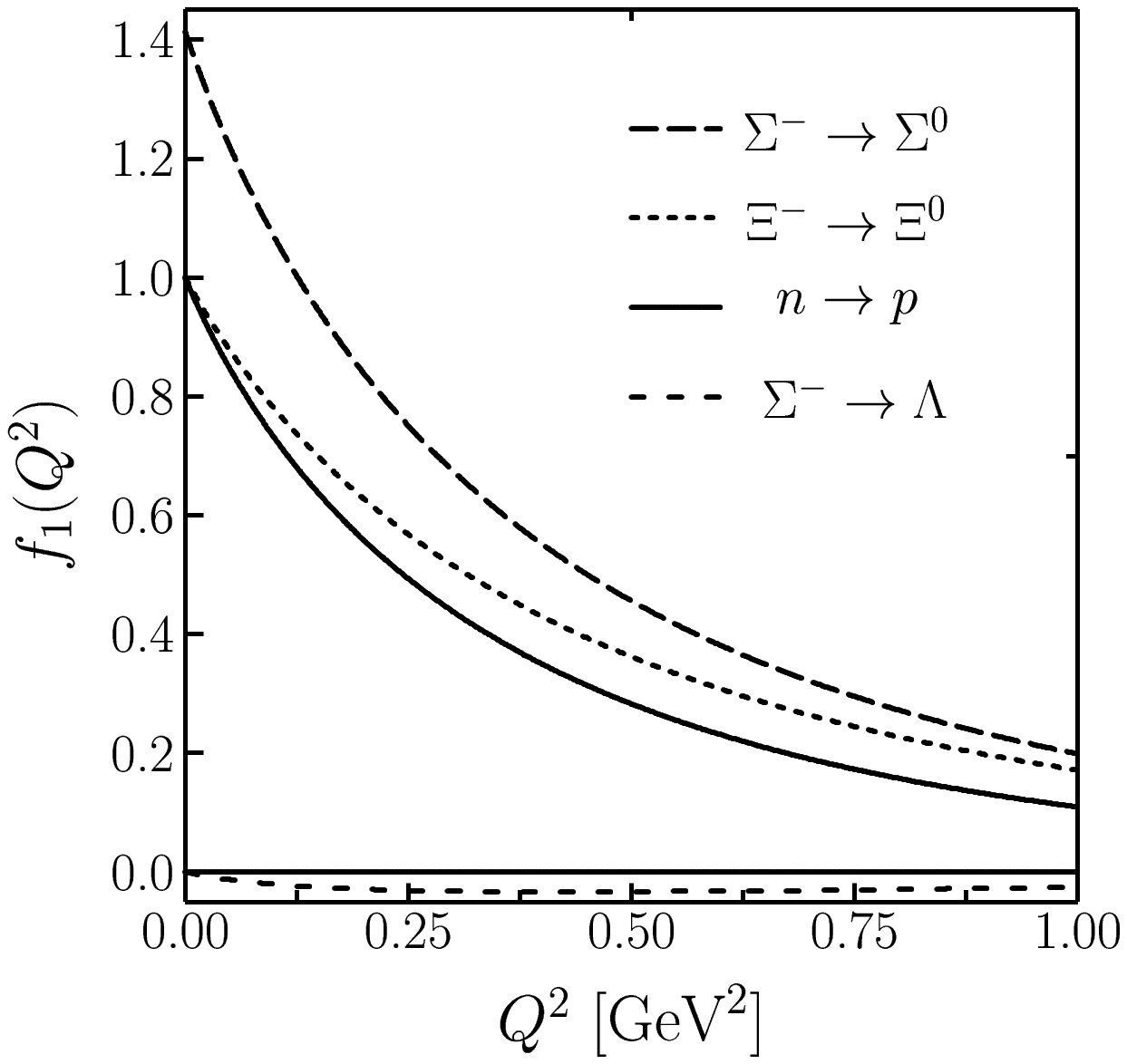}
~\includegraphics[scale=0.5]{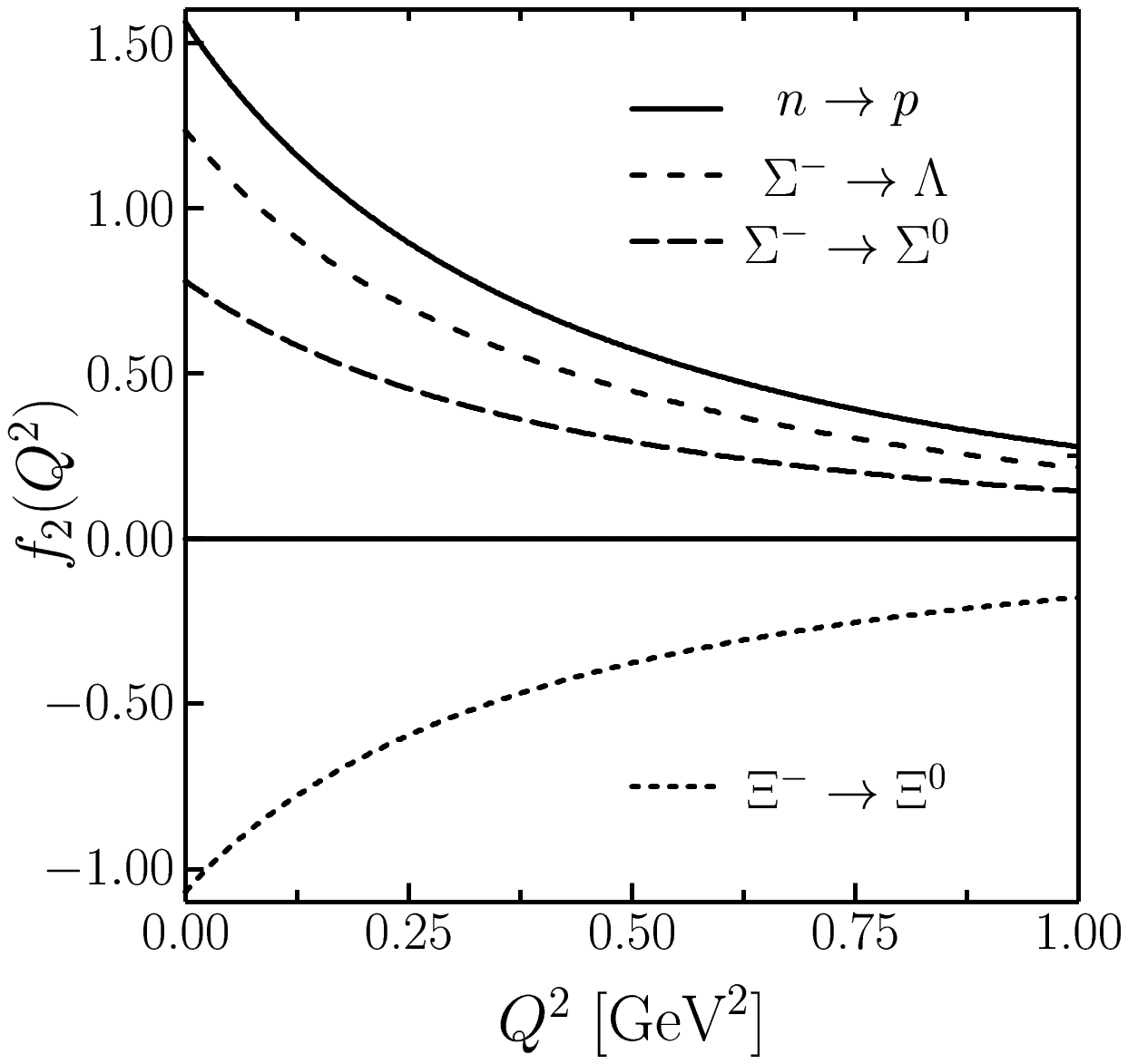}\vspace{0.3cm}

\includegraphics[scale=0.5]{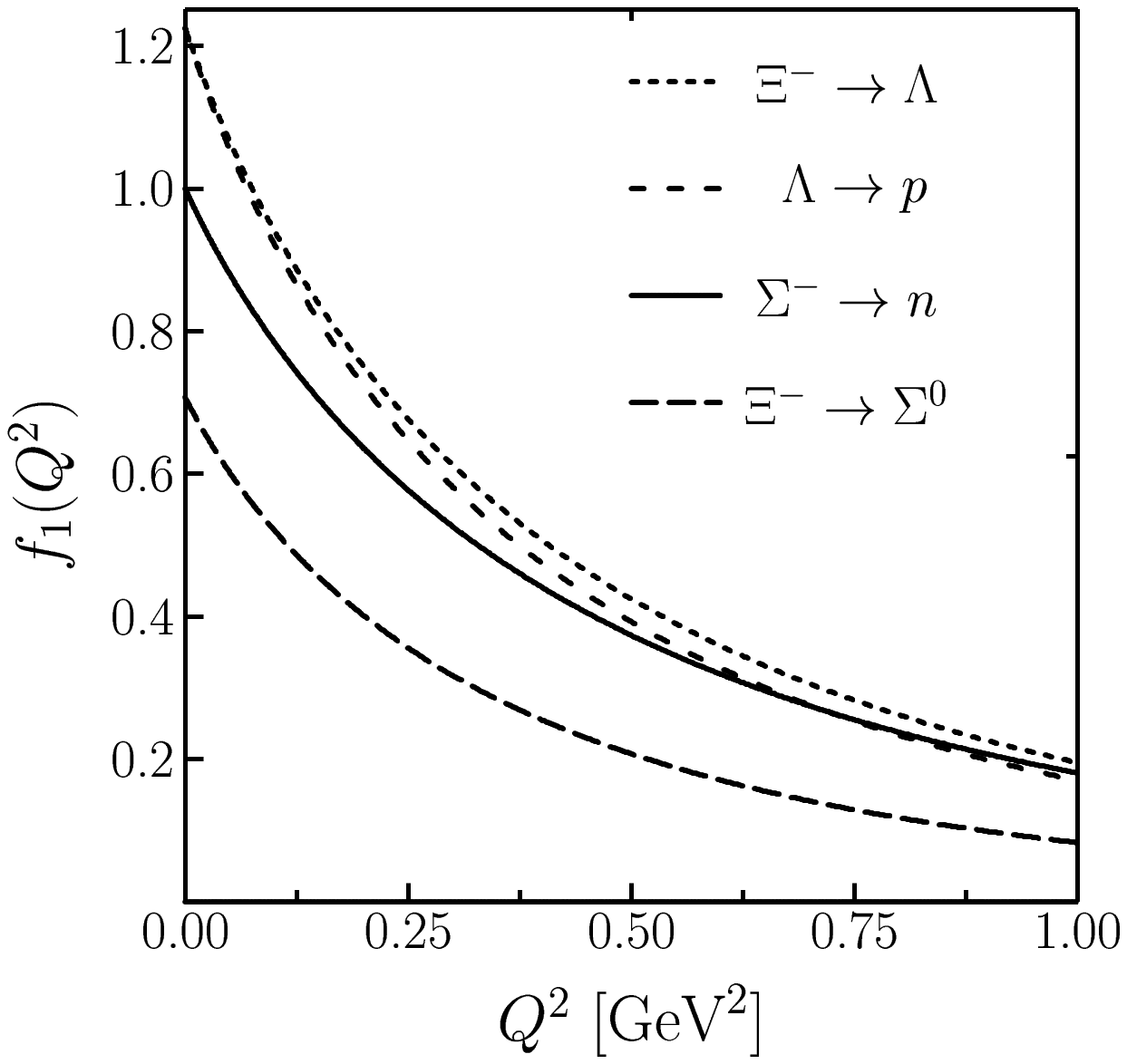}
~\includegraphics[scale=0.5]{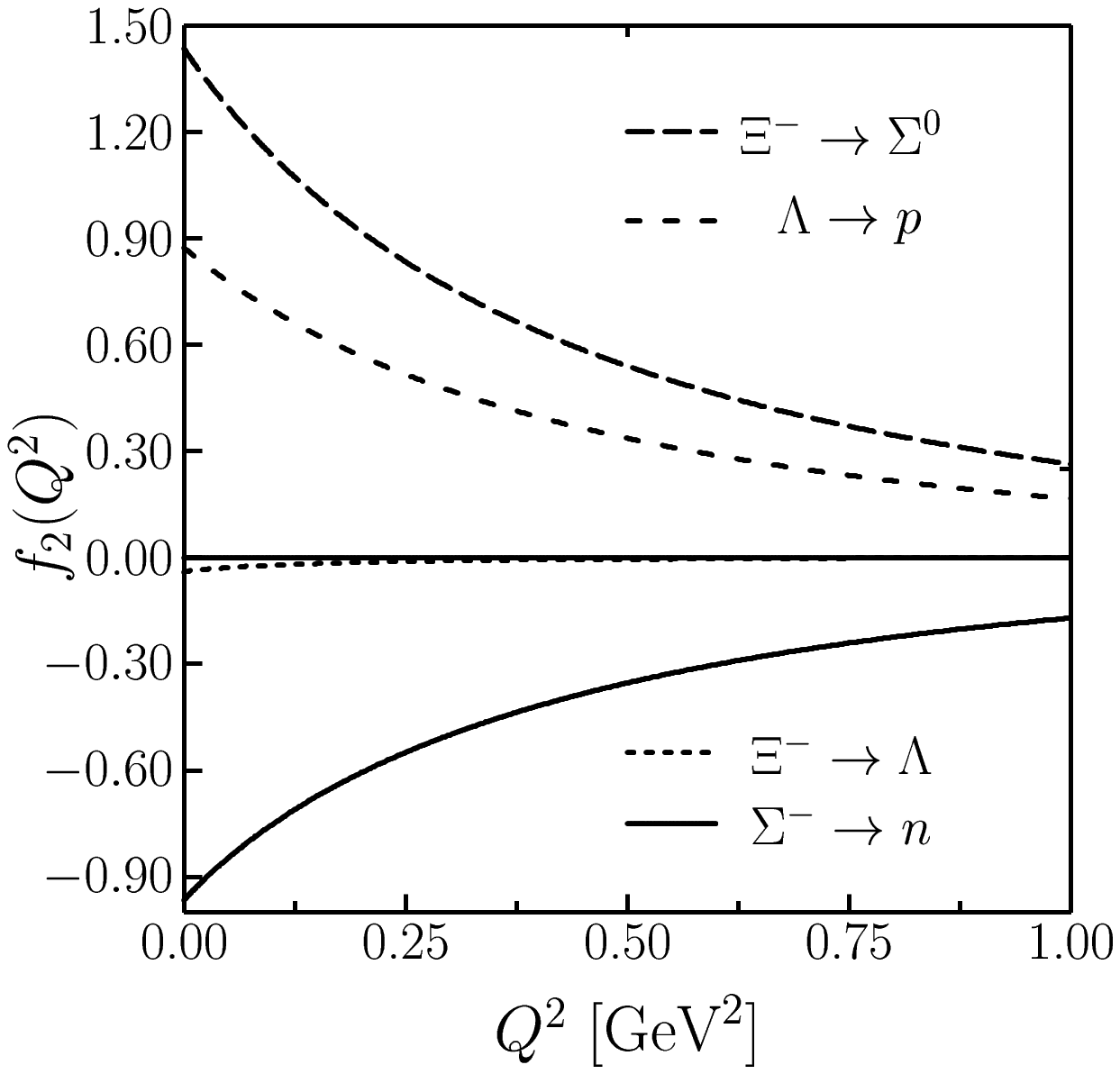}
\end{center}
\caption{\label{Transition f_1 f_2} SHD form factors
$f_{1}^{B_{1}B_{2}}(Q^{2})$ and $f_{2}^{B_{1}B_{2}}(Q^{2})$ in the
self-consitent $\chi$QSM for $M=420$ MeV with linear
$m_{s}$ corrections taken into account. In the upper panel, the form
factors for the $\Delta S=0$ transitions are drawn and in the lower
panel those for the $\Delta s=1$ transitions are depicted.}
\end{figure}

We now consider the SHD form factors for finite $Q^{2}$ up to
$Q^{2}\leq1\textrm{GeV}^{2}$ by starting with
$f_{1}^{B_{1}B_{2}}(Q^{2})$.   The $f_1$ form factors are depicted in
the left panel of Fig.\ref{Transition f_1 f_2}.  In general, the form
factors $f_{1}^{B_{1}B_{2}}(Q^{2})$ do not show any significant
dependence on the constituent quark mass $M$ in the range of $M=400$
MeV to $M=450$ MeV. Due to the Ademollo-Gatto theorem, the linear
$m_{s}$ corrections vanish exactly at $Q^{2}=0$.  For finite $Q^{2}$,
linear $m_{s}$ corrections are negligible for all transitions.  Only
for the $\Sigma^{-}\to\Lambda$ 
transition, they contribute to the form factor by about $10\,\%$.  We
fitted the $\chi$QSM results of the $f_{1}^{B_{1}B_{2}}(Q^{2})$ form
factors, using a dipole-type parameterization:
\begin{equation}
f_{1}^{B_{1}B_{2}}(Q^{2}) =
\frac{f_{1}^{B_{1}B_{2}}(0)}{(1+Q^{2}/M_{f_{1}}^{2})^{2}},
\end{equation}
where the radius of $f_{1}^{B_{1}B_{2}}(0)$ is given by $\langle
r_{f_{1}}\rangle^{2}=12/M_{f_{1}}^{2}\,\mathrm{fm}^2$.  The dipole
masses $M_{f_{1}}$ are listed in Table~\ref{Dipole-masses} for
$M=420$ MeV with linear $m_{s}$
corrections. Reference~\cite{Cabibbo2003} gives a dipole mass of 
$M_{f_{1}}^{np}=0.84\pm0.04$ GeV, while the present result is
$M_{f_{1}}^{np}=0.752$ GeV.  The dipole mass of the electric proton
form factor within the same scheme as used in the present work is
$M_{G_{E}^{p}}^{\chi\mathrm{QSM}}=0.779$ GeV whereas the empirical
dipole mass is $M_{G_{E}^{p}}^{\mathrm{exp}}\approx0.710$ GeV. Thus,
the dipole mass of the $n\to p$ SHD form factor is very similar to
that of the proton electric one.

As for the SHD $G^{B_1B_2}_M(Q^2)$ form factors, the linear $m_{s}$
corrections turn out to be small. In particular they are almost
negligible for the $\Delta S=0$ and contribute to the $\Delta S=1$
transitions by about $(5\sim11)\,\%$. This is due to the fact that
each term in the linear $m_s$ contributions interferes each other
destructively. The dependence on the constituent quark mass $M$ is
also very weak.  The form factors $G_{M}$ and $f_{2}$ can be also
parameterized in the dipole form as follows:
\begin{equation}
G_{M}^{B_{1}B_{2}}(Q^{2}) =
\frac{G_{M}^{B_{1}B_{2}}(0)}{(1+Q^{2}/M_{G_{M}}^{2})^{2}}, \;\;\;
f_{2}^{B_{1}B_{2}}(Q^{2})=\frac{f_{2}^{B_{1}B_{2}}(0)}{
  (1+Q^{2}/M_{f_{2}}^{2})^{2}}.
\end{equation}
The results of the corresponding dipole masses are listed in
Table~\ref{Dipole-masses} for $M=420$ MeV.  The radii are given by
$\langle r_{G_{M}}\rangle^{2}=12/M_{G_{M}}^{2}$ and $\langle
r_{f_{2}}\rangle^{2}=12/M_{f_{2}}^{2}$. The full form
factors are presented in the lower panel of Fig.\ref{Transition f_1 f_2}.

\begin{table}[ht]
\caption{\label{Dipole-masses}Dipole masses and radii of the SHD form
factors $f_1$, $G_M$, and $f_2$ in the self-consistent $\chi$QSM for 
$M=420$ MeV with linear $m_{s}$ corrections taken into account,
i.e. $\chi \textrm{QSM}_{m_{s}^{0}+m_{s}^{1}}$.\protect \\
}
\begin{tabular}{c|cc|cc|cc}
\hline
&
$M_{f_{1}}\,[\textrm{MeV}]$&
$\langle r_{f_{1}}^{2}\rangle\,[\textrm{fm}^{2}]$&
$M_{G_{M}}\,[\textrm{MeV}]$&
$\langle r_{G_{M}}^{2}\rangle\,[\textrm{fm}^{2}]$&
$M_{f_{2}}\,[\textrm{MeV}]$&
$\langle r_{f_{2}}^{2}\rangle\,[\textrm{fm}^{2}]$ \tabularnewline
\hline
$n\to p$&
$752$&
$0.826$&
$842$&
$0.659$&
$871$&
$0.616$\tabularnewline
$\Sigma^{-}\to\Lambda$&
$\cdots$&
$\cdots$&
$844$&
$0.656$&
$862$&
$0.629$\tabularnewline
$\Sigma^{-}\to\Sigma^{0}$&
$807$&
$0.717$&
$847$&
$0.651$&
$887$&
$0.594$\tabularnewline
$\Xi^{-}\to\Xi^{0}$&
$865$&
$0.624$&
$832$&
$0.675$&
$844$&
$0.656$\tabularnewline
\hline
$\Sigma^{-}\to n$&
$882$&
$0.600$&
$860$&
$0.632$&
$874$&
$0.612$\tabularnewline
$\Lambda\to p$&
$804$&
$0.723$&
$857$&
$0.636$&
$896$&
$0.582$\tabularnewline
$\Xi^{-}\to\Sigma^{0}$&
$766$&
$0.796$&
$862$&
$0.629$&
$889$&
$0.591$\tabularnewline
$\Xi^{-}\to\Lambda$&
$842$&
$0.659$&
$865$&
$0.624$&
$506$&
$1.824$\tabularnewline
\hline
\end{tabular}
\end{table}

It is also of great interest to compare the present results with the
recent investigation of $f_{2}^{B_{1}B_{2}}(0)$ and $\langle
r_{f_1}\rangle^2 $ for the $\Delta S=1$ transitions in $\chi$PT to
order $\mathcal{O}(p^4)$~\cite{CHPT_hyperon2007}.  As shown in
Table~\ref{CHPT_comp}, they are in good agreement each other except
for the $\Xi^{-}\to\Sigma^{0}$ transition. 

\begin{table}[ht]
\caption{\label{CHPT_comp} Comparison of the final results of $\chi
  \textrm{QSM}_{m_{s}^{0}+m_{s}^{1}}$ for $\langle
  r_{f_{1}}\rangle^{2}$ and $f_{2}(0)$ with those of $\chi$PT to
  order $\mathcal{O}(p^4)$~\cite{CHPT_hyperon2007} in which the 
  normalization is given as
  $f_{2}(0)=\kappa\frac{m_{1}}{2m_{p}}$.\protect \\ 
}
\begin{tabular}{c|cc}
\hline
$\langle r_{f_{1}}\rangle^{2}/\textrm{fm}^{2}$&
$\chi$PT $\mathcal{O}(p^4)$&
$\chi$QSM\tabularnewline
\hline
$\Sigma^{-}\to n$&
$0.72$&
$0.60$\tabularnewline
$\Lambda\to p$&
$0.70$&
$0.72$\tabularnewline
$\Xi^{-}\to\Sigma^{0}$&
$0.77$&
$0.80$\tabularnewline
$\Xi^{-}\to\Lambda$&
$0.65$&
$0.66$\tabularnewline
\hline
\end{tabular}~~\begin{tabular}{c|cc}
\hline
$f_{2}^{B_{1}B_{2}}(0)$&
$\chi$PT $\mathcal{O}(p^4)$&
$\chi$QSM\tabularnewline
\hline
$\Sigma^{-}\to n$&
$-1.00$&
$-0.96$\tabularnewline
$\Lambda\to p$&
$0.88$&
$0.88$\tabularnewline
$\Xi^{-}\to\Sigma^{0}$&
$1.90$&
$1.43$\tabularnewline
$\Xi^{-}\to\Lambda$&
$-0.05$&
$-0.03$\tabularnewline
\hline
\end{tabular}
\end{table}

\subsection{SHD axial-vector form factors}
\begin{figure}
\begin{center}
\includegraphics[scale=0.5]{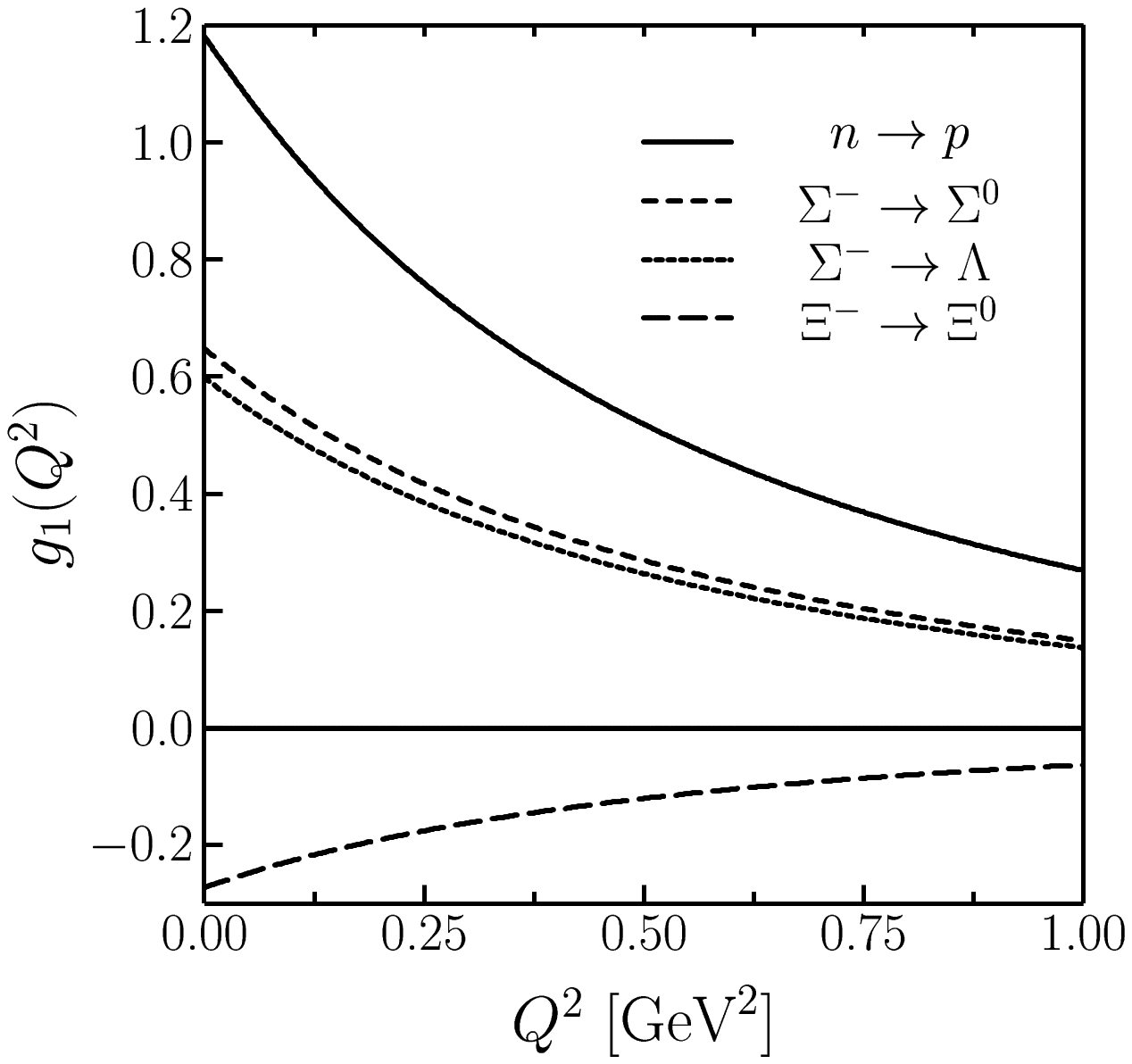}
~\includegraphics[scale=0.5]{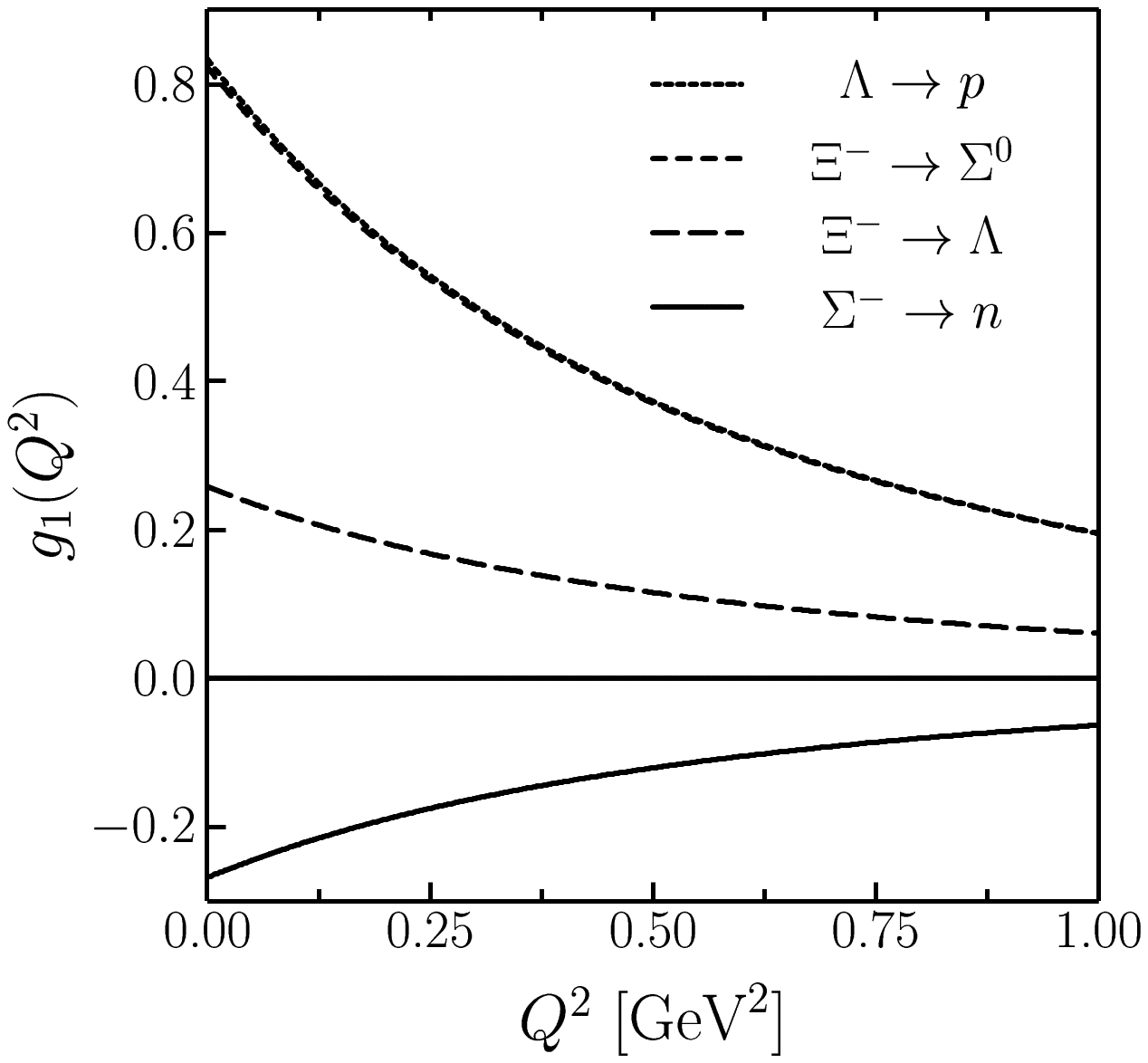}
\end{center}
\caption{\label{cap:Octet GAtrans}SHD axial-vector form factors
$g^{B_1 B_2}_1(Q^2)$ in the self-consistent $\chi$QSM for
$M=420$ MeV with linear $m_s$ corrections taken into account. In the
left and right panels, the form factors for the $\Delta S=0$ and
$\Delta S=1$ transitions are drawn, respectively.}
\end{figure}

We present now the results of the SHD axial-vector form factors
$g_{1}^{B_{1}B_{2}}(Q^{2})$. The SHD axial-vector constants are
insensitive to the constituent quark mass $M$.  Varying $M$ from $400$
to $450$ MeV, they are changed just by about $5\,\%$. The linear
$m_{s}$ corrections turn out to be also small.
Figure~\ref{cap:Octet GAtrans} depicts the SHD axial-vector form
factors $g_{1}^{B_{1}B_{2}}(Q^{2})$ for all relevant processes.  The
dipole parameterization is used for fitting the axial-vector
form factors in the $\chi$QSM:
\begin{equation}
g_{1}^{B_{1}B_{2}}(Q^{2}) =
\frac{g_{1}^{B_{1}B_{2}}(0)}{(1+Q^{2}/M_{g_{1}}^{2})^{2}},
\end{equation}
where the corresponding radius is again given as $\langle
r_{g_{1}}\rangle^{2}=12/M_{g_{1}}^{2}$.  The axial-vector constants
and dipole masses are given in Table~\ref{table-axial}.  In flavor
$SU(3)$ symmetry, the axial-vector part of the SHD can be
characterized by two form factors $F(Q^{2})$ and $D(Q^{2})$ that are
related to the proton triplet and octet axial-vector form factors:
\begin{eqnarray}
g_{A}^{3}(Q^{2}) & = & F(Q^{2})+D(Q^{2}),\cr
g_{A}^{8}(Q^{2}) & = &
\frac{1}{\sqrt{3}}\Big(3F(Q^{2})-D(Q^{2})\Big).
\end{eqnarray}
The triplet and octet axial-vector constants $g_{A}^{3}(0)$ and
$g_{A}^{8}(0)$ have been already calculated explicitly in the
$\chi$QSM~\cite{SilvaKUG:2005} by using exactly the present
formalism. The corresponding values of $F$ and $D$ are given below,
where linear $m_s$ corrections are taken into account: 
\begin{eqnarray}
g_{A}^{3}(0)=1.16, & \;\;\; &
F(0)=0.446,\;\;\;\;\; F(0)^{exp}=0.462\pm0.008,\cr
g_{A}^{8}(0)=0.36, &  & D(0)=0.714,\;\;\;\;\;
D(0)^{exp}=0.804\pm0.008.
\end{eqnarray}

In Table~\ref{cap:RatiosG/F}, the final results of the ratios
$(g_{1}(0)/f_{1}(0))^{B_{1}B_{2}}$ are listed with and without linear
$m_{s}$-corrections considered and are compared with the experimental
data.  Note that these ratios were also investigated in the $\chi$QSM
without the symmetry-conserving quantization~\cite{CQSM_semilep1998}
and in the infinite momentum frame~\cite{lorce_semilep} in flavor $SU(3)$
symmetry. T he flavor $SU(3)$-symmetry breaking contributions are
negligible and the effect of the symmetry-conserving quantization is
moderate.  

\begin{table}
\caption{\label{table-axial}Results of the SHD axial-vector
  $g_{1}^{B_{1}B_{2}}(0)$ constants and dipole masses
  $M_{g_{1}}^{B_{1}B_{2}}$ in the self-consistent $\chi$QSM for
$M=420$ MeV.  The first two columns show the results
without and with $m_s$ corrections, respectively. The third column
shows those calculated in $SU(3)$ symmetry by using the results of
the proton $g_A^3$ and $g_A^8$ constants in the $\chi$QSM. The
dipole masses are given in units of GeV with linear
$m_{s}$ corrections taken into account.}
\begin{center}
\begin{tabular}{c|ccc|c}
\hline
$g_1(0)$&
$\chi \textrm{QSM}_{\mathrm{SU(3)}}$&
$\chi \textrm{QSM}_{m_{s}^{0}+m_{s}^{1}}$&
$\chi \textrm{QSM}_{\mathrm{SU(3)}}$&
$M_{g_{1}}/$GeV\tabularnewline
\hline
$n\to p$&
$1.16$&
$1.18$&
$F+D=1.16$&
$1.04$\tabularnewline
$\Sigma^{-}\to\Lambda$&
$0.58$&
$0.60$&
$\sqrt{\frac{2}{3}}D=0.58$&
$1.04$\tabularnewline
$\Sigma^{-}\to\Sigma^{0}$&
$0.63$&
$0.65$&
$\sqrt{2}F=0.63$&
$1.04$\tabularnewline
$\Xi^{-}\to\Xi^{0}$&
$-0.27$&
$-0.27$&
$F-D=-0.27$&
$1.05$\tabularnewline
\hline
$\Sigma^{-}\to n$&
$-0.27$&
$-0.27$&
$F-D=-0.27$&
$1.05$\tabularnewline
$\Lambda\to p$&
$0.83$&
$0.83$&
$\sqrt{\frac{3}{2}}(F+D/3)=0.84$&
$1.06$\tabularnewline
$\Xi^{-}\to\Sigma^{0}$&
$0.82$&
$0.82$&
$\sqrt{\frac{1}{2}}(F+D)=0.82$&
$1.06$\tabularnewline
$\Xi^{-}\to\Lambda$&
$0.25$&
$0.26$&
$\sqrt{\frac{3}{2}}(F-D/3)=0.25$&
$1.05$\tabularnewline
\hline
\end{tabular}
\end{center}
\end{table}

\begin{table}
\caption{\label{cap:RatiosG/F}Results for the ratios
$(g_{1}/f_{1})$. The first two columns show the results
of this work with and without $m_s$ corrections, 
where our final results are given by $\chi
\textrm{QSM}_{m_{s}^{0}+m_{s}^{1}}$. The other two columns 
correspond to those of the $\chi$QSM in the Fock representation on 
the light cone \cite{lorce_semilep} and to those of the old 
$\chi$QSM~\cite{CQSM_semilep1998} without symmetry conserving
quantization, respectively.  Experimental data are taken from  
Ref.~\cite{PDG:2006,BGHORS:1982,BGHORS:1983}.  For the
$\Sigma^{-}\to\Lambda$ decay the value of $\sqrt{\frac{3}{2}}\cdot
g_{1}^{\Sigma\Lambda}(0)$ is listed.}
\begin{tabular}{c|cc|ccc}
\hline
$g_{1}/f_{1}$&
$\chi \textrm{QSM}_{\mathrm{SU(3)}}$&
$\chi \textrm{QSM}_{m_{s}^{0}+m_{s}^{1}}$&
IMF \cite{lorce_semilep}&
'98 \cite{CQSM_semilep1998}&
Exp.\tabularnewline
\hline
$n\to p$&
$1.16$&
$1.18$&
$1.156$&
$1.42$&
$1.2695\pm0.0029$\tabularnewline
$\Sigma^{-}\to\Lambda$&
$0.71$&
$0.73$&
$0.686$&
$0.91$&
$0.720\pm0.020$\tabularnewline
$\Sigma^{-}\to\Sigma^{0}$&
$0.45$&
$0.46$&
$0.470$&
$0.55$&
\tabularnewline
$\Xi^{-}\to\Xi^{0}$&
$-0.27$&
$-0.27$&
$-0.215$&
\tabularnewline
\hline
$\Sigma^{-}\to n$&
$-0.27$&
$-0.27$&
$-0.215$&
$-0.31$&
$-0.340\pm0.017$\tabularnewline
$\Lambda\to p$&
$0.68$&
$0.68$&
$0.699$&
$0.73$&
$0.718\pm0.015$\tabularnewline
$\Xi^{-}\to\Sigma^{0}$&
$1.16$&
$1.16$&
$1.156$&
$1.29$&
$1.25_{-0.16}^{+0.14}$\tabularnewline
$\Xi^{-}\to\Lambda$&
$0.20$&
$0.21$&
$0.242$&
$0.22$&
$0.25\pm0.05$\tabularnewline
\hline
\end{tabular}
\end{table}
\setcounter{table}{7}

It is also interesting to consider the ratio of $\Lambda\to
pe^{-}\overline{\nu}$ to $\Sigma^{-}\to ne^{-}\overline{\nu}$, since
it is known experimentally.  The comparison can be found in the
following:
\begin{equation}
\frac{(g_{1}/f_{1})[\Lambda\to
  pe^{-}\overline{\nu}]}{(g_{1}/f_{1})[\Sigma^{-}\to
  ne^{-}\overline{\nu}]}=-2.52,\;\;\;\;\textrm{Exp.:}-2.11\pm0.15.
\end{equation}
The result is in qualitative agreement with the data.

In flavor $SU(3)$ symmetry, the transitions
$\Xi^{0}\to\Sigma^{+}e^{-}\bar{\nu}_{e}$ and $n\to
pe^{-}\bar{\nu}_{e}$ are identical except that the valence
d quarks are replaced by s quarks in the initial and final state
baryons, respectively.  In this case, only the CKM matrix elements
become important to distinguish these decay amplitudes.  Experimental
values for the $\Xi^{0}\to\Sigma^{+}e^{-}\bar{\nu}_{e}$ decay are
available from Ref.~\cite{KTeV2001} and recently also from
Ref.\cite{NA48/I:2007}.  The extracted value of $g_{1}/f_{1}$ for
$\Xi^{0}\to\Sigma^{+}e^{-}\overline{\nu_{e}}$ is presented as
\begin{equation}
\Big{(} g_{1}/f_{1} \Big{)}^{\Xi^{0}\Sigma^{+}}_{\textrm{KTeV}} =
1.32_{-0.17\textrm{stat}}^{+0.21}\pm0.05_{\textrm{syst}}, \;\;\;\;
\Big{(} g_{1}/f_{1} \Big{)}^{\Xi^{0} \Sigma^{+}}_{\textrm{NA48/I}} =
1.20\pm0.05.
\end{equation}
In the present work with $SU(3)$-symmetry breaking, we obtain the
following result:
\begin{equation}
\Big(g_{1}/f_{1}\Big)^{\Xi^{0}\Sigma^{+}}_{\chi
  \textrm{QSM}}=\Big(\sqrt{2}\cdot
g_{1}/(\sqrt{2}\cdot
f_{1})\Big)^{\Xi^{-}\Sigma^{0}}_{\chi \textrm{QSM}}=\Big(\sqrt{2}\cdot
g_{1}/1\Big)^{\Xi^{-}\Sigma^{0}}_{\chi \textrm{QSM}}=1.16,
\end{equation}
which is close to the above experimental data as well as to the values
of $g_{1}/f_{1}$ for $n\to pe^{-}\bar{\nu}_{e}$
\begin{equation}
\Big{(}g_{1}/f_{1}\Big{)}^{np}_{\textrm{exp.}}=1.2695\pm0.0029,
\,\,\,\,\,\,\,\,\,\,\,\,\,\,\
\Big{(} g_{1} /f_{1} \Big{)}^{np}_{\chi \textrm{QSM}}=1.18,
\end{equation}
showing explicitly that $m_s$ contributions to the SHD are small.

In Ref.~\cite{HyperonV(us)2}, the CKM matrix element $V_{us}$ was
extracted within a large $N_c$ approach.  The SHD form factors $f_1$,
$f_2$ and $g_1$ were fitted to experimental decay rates, angular
correlation, and asymmetry coefficients.  The present work is
consistent with the values presented in Ref.~\cite{HyperonV(us)2}.
Very recently, a lattice study of the $\Sigma^{-}\to nl\nu$ decay was
performed~\cite{Lattice_Sigm2n}.  A polynomial extrapolation to the
physical point yields the value of $g_{1}(0)/f_{1}(0) =
-0.287\pm0.052$ with $g_2(0)=0$.  The corresponding result of the
present work in Table~\ref{cap:RatiosG/F} is $g_{1}(0)/f_{1}(0) =
-0.27$, which is consistent with the lattice one. The authors of
Ref.~\cite{Lattice_Sigm2n} also computed $g_2(0)$ and obtained a
rather large value of $g_2(0)/f_1(0)=+0.63 \pm 0.26$. They calculated 
 $|(g_1 - 0.133g_2)/f_1| = 0.37\pm 0.08$ which is in good agreement
with the experimental value of $|(g_1 - 0.133g_2)/f_1| =
0.327\pm0.007_{\textrm{stat}}\pm0.019_{\textrm{syst}}$ extracted
in Ref.~\cite{Sigm2n1988}.  The $g_2(Q^2)$ could also be calculated in
the $\chi$QSM and in this way one could check if also there a scenario
with a large $g_2(0)$ comes out.  In
Ref.~\cite{BetaDecayRelQuarkModel}, the SHD were 
investigated in a relativistic quark model, the results of the present
work are compatible with those of Ref.~\cite{BetaDecayRelQuarkModel}.
The dipole mass was extracted in
Refs.~\cite{Cabibbo2003,Gaillard1984}: $M_A=1.08\pm0.08$ GeV for the $n
\to p$ decay and $M_A=1.25$ GeV for $\Delta S=1$ decays.
References~\cite{Mendieta1998,Garcia1985} presented $M_A = 0.96$ GeV
for the $\Delta S =0$ decays and $M_A=1.11$ GeV for $\Delta S=1$. The
corresponding results of the present work are given as $M_A \approx
1.04 $ GeV and $M_A = 1.06 $ GeV.
\section{Summary and Conclusion}
In the present work we investigated the semileptonic hyperon decays
(SHD) within the framework of the self-consistent SU(3) chiral
quark-soliton model.  We take into account the linear 
rotational $1/N_c$ corrections as well as linear $m_s$ corrections and
employ the symmetry-conserving quantization.  In particular, we
calculated the SHD vector form factors $f_1(Q^2)$ and $f_2(Q^2)$ and
axial-vector form factors $g_1(Q^2)$ for all relevant decays in the
baryon octet.  Since $f_3(Q^2)$ and $g_3(Q^2)$ are always multiplied by
a factor of $(m_e/M_B)^2$ in the transition amplitudes, we neglected 
them. The form factor $g_2(Q^2)$ was neglected, assuming that it is
very small.

The chiral quark-soliton model has been applied for many years
successfully to various obseervables in the hadronic and partonic
sector.  All numerical parameters of the present investigation are the
same as in previous works and were fixed by only four basic pion and
nucleon observables.  A self-consistent soliton profile 
is used in order to solve numerically for eigenvalues of the
$\chi$QSM single-particle Hamiltonian.  These eigenvalues were then
used for calculating every form factor of the present work.

We first discussed the results for the SHD constants at $Q^2=0$ of the
form factors $f_1$ and $f_2$, see Table~\ref{tab f1gmf2} for the final
results.  Since we only consider the linear $m_s$-corrections, we
have the flavor-$SU(3)$ symmetric results for $f_1(0)$, which is a
consequence of the Ademollo-Gatto theorem.  Our values of $f_2/f_1$
agree qualitatively with those of the Cabibbo model without and with
$SU(3)$-symmetry breaking as well as with experimental data, see
Table~\ref{resf2/f1} for the final results.  Especially, the SHD
constants for the $\Sigma^- \to n$ and $\Xi^0 \to \Sigma^+$ processes
agree very well with the new experimental data.  The $m_s$ corrections
to $f_2(0)$ for both the $\Delta S=0$ and $\Delta S=1$ transitions
turn out to be sizeable in accrodance with previous calculations in
the $\chi$QSM.  The vector radii and dipole masses were listed in
Table~\ref{Dipole-masses}.  The dipole mass for the $n\to p$ decay was
obtained as $M^{np}_{f_1}=0.752$ GeV which is comparable to those
in the literature.  The $\Delta S=1$ constants $f_2(0)$ and radii of
$f_1(Q^2)$ in the $\chi$QSM do also agree with a calculation in chiral
perturbation theory to order $\mathcal{O}(p^4)$.

Second, we discussed the SHD constants and dipole masses of the
axial-vector form factor $g_1(Q^2)$, see Tables~\ref{table-axial} and
\ref{cap:RatiosG/F} for the final results.  The dipole masses for the
$\Delta S=0$ decays turned out to be $M_{g_1}=1.04$ GeV and for the
$\Delta S=1$ decays $M_{g_1}=1.06$ GeV.  The results of the SHD
axial-vector constants of the present work are in good agreement with 
calculations done in the Fock state representation of the $\chi$QSM on
the light cone.  Various $m_s$ corrections in the form factor
$g_1(Q^2)$ destructively interfere, so that the total $m_s$
corrections become rather small.  The application of the
symmetry-conserving quantization reduces the $m_s$ corrections even
further. The first determination of $(g_1/f_1)$ for the process
$\Sigma^- \to nl\nu$ on the lattice yielded a value of
$-0.287\pm0.052$, while the present work gives $-0.27$.  In addition,
the chiral quark-soliton model reproduces very well the recent  
measured value of $(g_1/f_1)=1.20\pm0.05$ by the NA48/I
collaboration.  Overall, the chiral quark-soliton model with the
present techniques reproduces the existing experimental data very
accurately. Since the chiral quark-soliton model is the simplest
current quark model showing spontaneous breaking of chiral symmetry
the present calculation shows how important this effect is for the
physics of light baryons.   

\section*{Acknowledgments}
The work is supported by the DFG-Transregio-Sonderforschungsbereich
Bonn-Bochum-Giessen, the Verbundforschung (Hadrons and Nuclei) of the
Federal Ministry for Education and Research (BMBF) of Germany, the
Graduiertenkolleg Bochum-Dortmund, the COSY-project J\"ulich as well
as the EU Integrated Infrastructure Initiative Hadron Physics Project
under contract number RII3-CT-2004-506078.
The present work is also supported by the Korea Research Foundation
Grant funded by the Korean Government(MOEHRD) (KRF-2006-312-C00507).
T. Ledwig was also supported by a DAAD doctoral exchange-scholarship.
K.G. acknowledges the hospitality of the ECT* at Trento (Italy). 
\begin{appendix}
\section{Form Factors in the $\chi$QSM\label{app:a}}
The electric density in Eq.(\ref{eq:G_E cqsm}) is given by
\begin{eqnarray}
\mathcal{G}_{E}^{\chi}({\bm z}) & = &
D_{\chi8}^{(8)}\sqrt{\frac{1}{3}}\mathcal{B}({\bm
  z})-\frac{2}{I_{1}}D_{\chi i}^{(8)}J_{i}\mathcal{I}_{1}({\bm
  z})-\frac{2}{I_{2}}D_{\chi a}^{(8)}J_{a}\mathcal{I}_{2}({\bm
  z})\nonumber \\
 &  & -\frac{2}{\sqrt{3}}M_{1}D_{\chi8}^{(8)}\mathcal{C}({\bm
   z})-\frac{2}{3}M_{8}D_{88}^{(8)}D_{\chi8}^{(8)}\mathcal{C}({\bm
   z})\nonumber \\
 &  & +4\frac{K_{1}}{I_{1}}M_{8}D_{8i}^{(8)}D_{\chi i}^{(8)}({\bm
   z})\mathcal{I}_{1}({\bm
   z})+4\frac{K_{2}}{I_{2}}M_{8}D_{8a}^{(8)}D_{\chi
   a}^{(8)}\mathcal{I}_{2}({\bm z})\nonumber \\
 &  & -4M_{8}D_{8i}^{(8)}D_{\chi
   i}^{(8)}\mathcal{K}_{1}({\bm z})-4M_{8}D_{\chi
   a}^{(8)}D_{8a}^{(8)}\mathcal{K}_{2}({\bm z}),
\label{eq:Ele dens}
\end{eqnarray}
and the magnetic density in Eq.(\ref{eq:G_M cqsm}) by
\begin{eqnarray}
\mathcal{G}_{M}^{\chi}({\bm z}) & = &
-\sqrt{3}D_{\chi3}^{(8)}\mathcal{Q}_{0}({\bm
  z})-\frac{1}{\sqrt{3}}\frac{1}{I_{1}}D_{\chi8}^{(8)}J_{3}\mathcal{X}_{1}({\bm
  z})+\sqrt{3}\frac{1}{I_{1}}d_{ab3}D_{\chi
  b}^{(8)}J_{a}\mathcal{X}_{2}({\bm
  z})+\sqrt{\frac{1}{2}}\frac{1}{I_{1}}D_{\chi3}^{(8)}\mathcal{Q}_{1}({\bm
  z})\nonumber \\
 &  &
 +\frac{2}{\sqrt{3}}\frac{K_{1}}{I_{1}}M_{8}D_{83}^{(8)}D_{\chi8}^{(8)}
 \mathcal{X}_{1}({\bm
   z})-2\sqrt{3}\frac{K_{2}}{I_{2}}M_{8}D_{8a}^{(8)}D_{\chi
   b}^{(8)}d_{ab3}\mathcal{X}_{2}({\bm z})\nonumber \\
 &  &
 +2\sqrt{3}\Big[M_{1}D_{\chi3}^{(8)} +
 \frac{1}{\sqrt{3}}M_{8}D_{88}^{(8)}
 D_{\chi3}^{(8)}\Big]\mathcal{M}_{0}({\bm z})\nonumber \\
 &  &
 -\frac{2}{\sqrt{3}}M_{8}D_{83}^{(8)}D_{\chi8}^{(8)}\mathcal{M}_{1}({\bm
   z}) + 2\sqrt{3}M_{8}D_{\chi
   a}^{(8)}D_{8b}^{(8)}d_{ab3}\mathcal{M}_{2}({\bm z}).
\label{eq:Mag dens}
\end{eqnarray}
The axial-vector density in Eq.(\ref{eq:G_A cqsm}) is given by
\begin{eqnarray}
\mathcal{G}_{10}^{\chi}({\bm z}) & = &
-\sqrt{\frac{1}{3}}D_{\chi3}^{(8)}\mathcal{A}({\bm
  z})+\frac{1}{3\sqrt{3}}\frac{1}{I_{1}}D_{\chi8}^{(8)}J_{3}
\mathcal{B}({\bm z})-\sqrt{\frac{1}{3}}\frac{1}{I_{2}}D_{\chi
  a}^{(8)}J_{b}d^{ab3}\mathcal{C}({\bm z})\nonumber \\
 &  &
 -\frac{1}{3\sqrt{2}}\frac{1}{I_{1}}D_{\chi3}^{(8)}\mathcal{D}({\bm
   z})-\frac{2}{3\sqrt{3}}\frac{K_{1}}{I_{1}}M_{8}D_{83}^{(8)}
 D_{\chi8}^{(8)}\mathcal{B}({\bm
   z})+\frac{2}{\sqrt{3}}\frac{K_{2}}{I_{2}}\, M_{8}\,
 D_{8a}^{(8)}D_{\chi b}^{(8)}d^{ab3}\mathcal{C}({\bm z})\nonumber \\
 &  &
 -\frac{2}{\sqrt{3}}\Big[M_{1}D_{\chi3}^{(8)}+\frac{1}{\sqrt{3}}
 M_{8}D_{88}^{(8)}D_{\chi3}^{(8)}\Big]\mathcal{H}({\bm z})\nonumber \\
 &  &
 +\frac{2}{3\sqrt{3}}M_{8}D_{83}^{(8)}D_{\chi8}^{(8)}
\mathcal{I}({\bm z})-\frac{2}{\sqrt{3}}M_{8}D_{8a}^{(8)}
D_{\chi b}^{(8)}d^{ab3}\mathcal{J}({\bm z}).
\label{eq:Ax dens}
\end{eqnarray}
The explicit expressions for $\mathcal{B}({\bm
  z}),\mathcal{I}_{1}({\bm z}),\cdots,\mathcal{J}({\bm z})$
can be found in the following Appendices for each form factor.
The baryon matrix-elements, such as $\langle
B^{\prime}|D_{\chi3}^{(8)}|B\rangle$, are evaluated by using the
$SU(3)$ group algebra~\cite{CHP,deSwart}
\begin{eqnarray}
\langle B_{\mathcal{R}^{\prime}}^{\prime}|D_{\chi
  m}^{n}(A)|B_{\mathcal{R}}\rangle & = &
\sqrt{\frac{\textrm{dim}\mathcal{R}^{\prime}}{\textrm{dim}
    \mathcal{R}}}(-1)^{\frac{1}{2}Y_{s}^{\prime}+S_{3}^{\prime}}
(-1)^{\frac{1}{2}Y_{s}+S_{3}}\nonumber
\\
 &  & \times\sum_{\gamma}\left(\begin{array}{ccc}
\mathcal{R}^{\prime} & n & \mathcal{R}_{\gamma}\\
Q^{\prime} & \chi & Q\end{array}\right)\left(\begin{array}{ccc}
\mathcal{R}^{\prime} & n & \mathcal{R}_{\gamma}\\
-Y_{s}^{\prime}S^{\prime}-S_{3}^{\prime} & m &
-Y_{S}S-S_{3}\end{array}\right)\,\,\,,
\end{eqnarray}
with $Q=YII_{3}$.  $\left(\cdots\right)$ denote the $SU(3)$
Clebsch-Gordan coefficients.  The results for all transitions in
Eqs.(\ref{eq:DS0 processes},\ref{eq:DS1 processes}) are listed below.
\section{Form Factor Densities}
\subsection{$\chi$QSM Electric Densities}
The electric densities are
\begin{eqnarray*}
\frac{1}{N_{c}}\mathcal{B}({\bm z}) & = & \varphi_{v}^{\dagger}({\bm
  z})\varphi_{v}({\bm
  z})-\frac{1}{2}\sum_{n}\textrm{sign}(\varepsilon_{n})\phi_{n}^{\dagger}({\bm
  z})\phi_{n}({\bm z}),\cr
\frac{\delta^{ij}}{N_{c}}\mathcal{I}_{1}({\bm z}) & = &
\frac{1}{2}\sum_{\varepsilon_{n}\neq\varepsilon_{v}}\frac{1}{
\varepsilon_{n}-\varepsilon_{v}}\langle
v|\tau^{i}|n\rangle\phi_{n}^{\dagger}({\bm z})\tau^{j}\phi_{v}({\bm
  z})+\frac{1}{4}\sum_{n,m}\mathcal{R}_{3}(\varepsilon_{n},
\varepsilon_{m})\langle
n|\tau^{i}|m\rangle\phi_{m}^{\dagger}({\bm z})\tau^{j}\phi_{n}({\bm z}),\cr
\frac{1}{N_{c}}\mathcal{I}_{2}({\bm z}) & = &
\frac{1}{4}\sum_{\varepsilon_{n^{0}}}\frac{1}{\varepsilon_{n^{0}} -
  \varepsilon_{v}}\langle n^{0}|v\rangle\phi_{v}^{\dagger}({\bm
  z})\phi_{n^{0}}({\bm z}) + \frac{1}{4} \sum_{n,m^{0}}
\mathcal{R}_{3}(\varepsilon_{n},\varepsilon_{m^{0}})
\phi_{m^{0}}^{\dagger}({\bm z})\phi_{n}({\bm z})\langle
n|m^{0}\rangle,\cr
\frac{1}{N_{c}}\mathcal{C}({\bm z}) & = &
\sum_{\varepsilon_{n}\neq\varepsilon_{v}}\frac{1}{\varepsilon_{n} -
  \varepsilon_{v}}\phi_{v}^{\dagger}({\bm z}) \phi_{n}({\bm z})\langle
n|\gamma^{0}|v\rangle+\frac{1}{2}\sum_{n,m}\langle
n|\gamma^{0}|m\rangle\phi_{m}^{\dagger}({\bm z})\phi_{n}({\bm
  z})\mathcal{R}_{5}(\varepsilon_{n},\varepsilon_{m}),\cr
\frac{\delta^{ij}}{N_{c}}\mathcal{K}_{1}({\bm z}) & = & \frac{1}{2}
\sum_{\varepsilon_{n}\neq\varepsilon_{v}}\frac{1}{\varepsilon_{n} -
  \varepsilon_{v}}\langle
v|\gamma^{0}\tau^{i}|n\rangle\phi_{n}^{\dagger}({\bm
  z})\tau^{j}\phi_{v}({\bm z})+\frac{1}{4}\sum_{n,m}\langle
n|\gamma^{0}\tau^{i}|m\rangle\phi_{m}^{\dagger}({\bm
  z})\tau^{j}\phi_{n}({\bm
  z})\mathcal{R}_{5}(\varepsilon_{n},\varepsilon_{m}),\cr
\frac{1}{N_{c}}\mathcal{K}_{2}({\bm z}) & = & \frac{1}{4}
\sum_{\varepsilon_{n^{0}}}\frac{1}{\varepsilon_{n^{0}} -
  \varepsilon_{v}}\phi_{v}^{\dagger}({\bm z}) \phi_{n^{0}}({\bm
  z})\langle n^{0}|\gamma^{0}|v\rangle + \frac{1}{4} \sum_{n,m}
\mathcal{R}_{5}(\varepsilon_{n},\varepsilon_{m^{0}})
\phi_{m^{0}}^{\dagger}({\bm z})\phi_{n}({\bm z})\langle
n|\gamma^{0}|m^{0}\rangle.
\end{eqnarray*}
The vectors $\langle n|$ are eigenstates of the $\chi$QSM Hamiltonian
$H(U)$ which are a linear combination of the eigenstates $\langle n^{0}|$
of the Hamiltonian $H(1)$~\cite{WakamatsuBasis}.

\subsection{$\chi$QSM Magnetic Densities}
The operator for the magnetic form factors in the $\chi$QSM is
$O_{1}=\gamma^{0}[{\bm z}\times{\bm \gamma}]_{3}=\gamma^{5}[{\bm
  z}\times{\bm \sigma}]_{10}$ and the magnetic densities are
\begin{eqnarray*}
\frac{1}{N_{c}}\mathcal{Q}_{0}({\bm z}) & = & \langle v||{\bm
  z}\rangle\{ O_{1}\otimes\tau_{1}\}_{0}\langle{\bm
  z}||v\rangle+\sum_{n}\sqrt{2G_{n}+1}\langle n||{\bm z}\rangle\{
O_{1}\otimes\tau_{1}\}_{0}\langle{\bm
  z}||n\rangle\mathcal{R}_{1}(\varepsilon_{n}),\cr
\frac{1}{N_{c}}\mathcal{X}_{1}({\bm z}) & = &
\sum_{\varepsilon_{n}\neq\varepsilon_{v}}\frac{1}{\varepsilon_{n} -
  \varepsilon_{v}}(-)^{G_{n}}\langle
v||{\bm z}\rangle O_{1}\langle{\bm z}||n\rangle\langle
n||\tau_{1}||v\rangle\cr
 &  & +\frac{1}{2}\sum_{n,m}\mathcal{R}_{5}(\varepsilon_{n},
 \varepsilon_{m})(-)^{G_{m}-G_{n}} \langle n||\tau_{1}||m
 \rangle\langle m||{\bm z} \rangle O_{1}\langle{\bm z}|| n\rangle,\cr
\frac{1}{N_{c}}\mathcal{X}_{2}({\bm z}) & = &
\sum_{\varepsilon_{n^{0}}} \frac{1}{\varepsilon_{n^{0}} -
  \varepsilon_{v}} \langle n^{0}||{\bm z}\rangle\{
O_{1}\otimes\tau_{1}\}_{0} \langle{\bm z}||v\rangle \langle v\mid
n^{0}\rangle\cr
 &  &
 +\sum_{n,m^{0}}\mathcal{R}_{5}(\varepsilon_{n},\varepsilon_{m^{0}})
 \sqrt{2G_{m}+1} \langle m^{0}||{\bm z}\rangle\{
 O_{1}\otimes\tau_{1}\}_{0} \langle{\bm z}||n\rangle \langle n \mid
 m^{0}\rangle,\cr
\frac{1}{N_{c}}\mathcal{Q}_{1}({\bm z}) & = & \sum_{\varepsilon_{n}}
\frac{\textrm{sign}( \varepsilon_{n})}{\varepsilon_{n} -
  \varepsilon_{v}} (-)^{G_{n}}\langle n||{\bm z} \rangle\{
O_{1}\otimes \tau_{1}\}_{1}\langle{\bm z}||v\rangle\langle v||
\tau_{1}||n\rangle\cr
 &  & +\frac{1}{2}\sum_{n,m}\mathcal{R}_{4}(\varepsilon_{n},
 \varepsilon_{m})(-)^{G_{m}-G_{n}}\langle n||{\bm z}\rangle\{
 O_{1}\otimes\tau_{1}\}_{1}\langle{\bm z}||m\rangle\langle
 m||\tau_{1}||n\rangle,\cr
\frac{1}{N_{c}}\mathcal{M}_{0}({\bm z}) & = &
\sum_{\varepsilon_{n}\neq\varepsilon_{v}}\frac{1}{\varepsilon_{n} -
  \varepsilon_{v}}\langle
v||{\bm z}\rangle\{ O_{1}\otimes\tau_{1}\}_{0}\langle{\bm
  z}||n\rangle\langle n|\gamma^{0}|v\rangle\cr
 &  &
 -\frac{1}{2}\sum_{n,m}\mathcal{R}_{2}(\varepsilon_{n},\varepsilon_{m})
 \sqrt{2G_{m}+1}\langle
 n|\gamma^{0}|m\rangle\langle m||{\bm z}\rangle\{
 O_{1}\otimes\tau_{1}\}_{0}\langle{\bm z}||n\rangle,\cr
\frac{1}{N_{c}}\mathcal{M}_{1}({\bm z}) & = &
\sum_{\varepsilon_{n}\neq\varepsilon_{v}}
\frac{1}{\varepsilon_{n}-\varepsilon_{v}} (-)^{G_{n}}\langle n||
\gamma^{0}\tau_{1}|| v\rangle\langle v|| {\bm z}\rangle O_{1}
\langle{\bm z}|| n\rangle\cr
 &  & -\frac{1}{2}\sum_{n,m}\mathcal{R}_{2}(\varepsilon_{n},
 \varepsilon_{m}) (-)^{G_{m}-G_{n}}\langle n||\gamma^{0}\tau_{1}|| m
 \rangle \langle m||{\bm z}\rangle O_{1}\langle{\bm z}|| n\rangle,\cr
\frac{1}{N_{c}}\mathcal{M}_{2}({\bm z}) & = &
\sum_{\varepsilon_{n^{0}}} \frac{1}{\varepsilon_{n^{0}} -
  \varepsilon_{v}} \langle v||{\bm z}\rangle\{
O_{1}\otimes\tau_{1}\}_{0} \langle{\bm z}||n^{0}\rangle\langle
n^{0}|\gamma^{0}|v \rangle\cr
 &  &
 -\sum_{n,m^{0}}\mathcal{R}_{2}(\varepsilon_{n},\varepsilon_{m^{0}})
 \sqrt{2G_{m}+1} \langle m^{0}||{\bm z}\rangle\{
 O_{1}\otimes\tau_{1}\}_{0} \langle{\bm z}||n\rangle\langle
 n|\gamma^{0}|m^{0}\rangle.
\end{eqnarray*}

\subsection{$\chi$QSM Axial-vector Densities \label{app ax dens} }
Two parts in Eq.(\ref{eq:G_A cqsm}) correspond to the density
Eq.(\ref{eq:Ax dens}) with the expressions
$\mathcal{A}({\bm z}),\mathcal{B}({\bm z})...$ once calculated with
the operator
\begin{equation}
O_{1}=\gamma^{0}\gamma_{3}\gamma^{5}\,\,\,\,\,\,\,\,\,\,\,\,\,\,\,\textrm{and
once with}\,\,\,\,\,\,\,\,\,\,\,\,\,\,\,
O_{1}=\gamma^{0}\{ Y_{2}\otimes\gamma_{1}\}_{10}\gamma^{5}=1\cdot\{
Y_{2}\otimes\gamma_{1}\}_{10}\,\,\, ,
\end{equation}
yielding once $\mathcal{G}_{10}^{\chi}({\bm z})$ and $\{
Y_{2}\otimes\mathcal{G}_{1}^{\chi}({\bm z})\}_{10}$ in the
tensor-notation of \cite{Quantum_angular_theory}, respectively.
We have for $\mathcal{A}({\bm z}),\mathcal{B}({\bm z})...$:
\begin{eqnarray*}
\frac{1}{N_{c}}\mathcal{A}({\bm z}) & = & \langle v||{\bm z}\rangle\{
O_{1}\otimes\tau_{1}\}_{0}\langle{\bm
  z}||v\rangle+\sum_{n}\sqrt{2G+1}\mathcal{R}_{1}(\varepsilon_{n})\langle
n||{\bm z}\rangle\{ O_{1}\otimes\tau_{1}\}_{0}\langle{\bm
  z}||n\rangle,\cr
\frac{1}{N_{c}}\mathcal{B}({\bm z}) & = &
\sum_{\varepsilon_{n}\neq\varepsilon_{v}}\frac{1}{\varepsilon_{v} -
  \varepsilon_{n}}(-)^{G_{n}} \langle
n||{\bm z}\rangle O_{1}\langle{\bm z}||v\rangle\langle
v||\tau_{1}||n\rangle\\
 &  & -\frac{1}{2}\sum_{n,m}(-)^{G_{n}-G_{m}}\langle m||{\bm z}\rangle
 O_{1}\langle{\bm z}||n\rangle\langle
 n||\tau_{1}||m\rangle\mathcal{R}_{5}(\varepsilon_{n},\varepsilon_{m}),\\
\frac{1}{N_{c}}\mathcal{C}({\bm z}) & = &
\sum_{\varepsilon_{n^{0}}}\frac{1}{\varepsilon_{v}-\varepsilon_{n^{0}}}\langle
v||{\bm z}\rangle\{ O_{1}\otimes\tau_{1}\}_{0}\langle{\bm
  z}||n^{0}\rangle\langle n^{0}|v\rangle\\
 &  & -\sum_{n,m}\sqrt{2G_{n}+1}\langle n||{\bm z}\rangle\{
 O_{1}\otimes\tau_{1}\}_{0}\langle{\bm z}||m^{0}\rangle\langle
 m^{0}|n\rangle\mathcal{R}_{5}(\varepsilon_{n},\varepsilon_{m^{0}}),\\
\frac{1}{N_{c}}\mathcal{D}({\bm z}) & = &
\sum_{n}\frac{\textrm{sign}(\varepsilon_{n})}{\varepsilon_{v} -
  \varepsilon_{n}}(-)^{G_{n}}\langle
n||{\bm z}\rangle\{ O_{1}\otimes\tau_{1}\}_{1}\langle{\bm
  z}||v\rangle\langle v||\tau_{1}||n\rangle\\
 &  &
 +\frac{1}{2}\,\sum_{n,m}\mathcal{R}_{4}(\varepsilon_{n},
 \varepsilon_{m})(-)^{G_{n}-G_{m}}\langle
 m||{\bm z}\rangle\{ O_{1}\otimes\tau_{1}\}_{1}\langle{\bm
   z}||n\rangle\langle n||\tau_{1}||m\rangle,\\
\frac{1}{N_{c}}\mathcal{H}({\bm z}) & = &
\sum_{\varepsilon_{n}\neq\varepsilon_{v}}\,\frac{1}{\varepsilon_{v}-\varepsilon_{n}}\langle
v||{\bm z}\rangle\{ O_{1}\otimes\tau_{1}\}_{0}\langle{\bm
  z}||n\rangle\langle n|\gamma^{0}|v\rangle\\
 &  &
 +\frac{1}{2}\sum_{n,m}\mathcal{R}_{2}(\varepsilon_{n},\varepsilon_{m})
 \sqrt{2G_{m}+1}\langle  m||{\bm z}\rangle\{
 O_{1}\otimes\tau_{1}\}_{0} \langle{\bm
   z}||n\rangle\langle n|\gamma^{0}|m\rangle,\\
\frac{1}{N_{c}}\mathcal{I}({\bm z}) & = &
\sum_{\varepsilon_{n}\neq\varepsilon_{v}}\,\frac{1}{\varepsilon_{v} -
  \varepsilon_{n}}(-)^{G_{n}}\langle
v||{\bm z}\rangle O_{1}\langle{\bm z}||n\rangle\langle
n||\gamma^{0}\tau_{1}||v\rangle\\
 &  &
 +\frac{1}{2}\sum_{n,m}\mathcal{R}_{2}(\varepsilon_{n},\varepsilon_{m})
 (-)^{G_{n}-G_{m}}\langle
 m||{\bm z}\rangle O_{1}\langle{\bm z}||n\rangle\langle
 n||\gamma^{0}\tau_{1}||m\rangle,\\
\frac{1}{N_{c}}\mathcal{J}({\bm z}) & = &
\sum_{\varepsilon_{n^{0}}}\,\frac{1}{\varepsilon_{v}-\varepsilon_{n^{0}}}\langle
v||{\bm z}\rangle\{ O_{1}\otimes\tau_{1}\}_{0}\langle{\bm
  z}||n^{0}\rangle\langle n^{0}|\gamma^{0}|v\rangle\\
 &  &
 +\sum_{n,m}\mathcal{R}_{2}(\varepsilon_{n^{0}},\varepsilon_{m})
 \sqrt{2G_{m}+1}\langle
 m||{\bm z}\rangle\{ O_{1}\otimes\tau_{1}\}_{0}\langle{\bm
   z}||n^{0}\rangle\langle n^{0}|\gamma^{0}|m\rangle.
\end{eqnarray*}

\subsection{Regularization Functions}
The regularization functions are defined as:
\begin{eqnarray}
\mathcal{R}_{1}(\varepsilon_{n}) & = &
-\frac{1}{2\sqrt{\pi}}\varepsilon_{n}\int_{1/\Lambda^{2}}^{\infty}
\frac{du}{\sqrt{u}}e^{-u\varepsilon_{n}^{2}},\cr
\mathcal{R}_{2}(\varepsilon_{n},\varepsilon_{m}) & = &
\int_{1/\Lambda^{2}}^{\infty}du\frac{1}{2\sqrt{\pi
    u}}\frac{\varepsilon_{m}e^{-u\varepsilon_{m}^{2}} -
  \varepsilon_{n}e^{-u\varepsilon_{n}^{2}}}{
  \varepsilon_{n}-\varepsilon_{m}},\cr
\mathcal{R}_{3}(\varepsilon_{n},\varepsilon_{m}) & = &
\frac{1}{2\sqrt{\pi}}\int_{1/\Lambda^{2}}^{\infty}\frac{du}{\sqrt{u}}
\Big[\frac{1}{u}
\frac{e^{-\varepsilon_{n}^{2}u}-e^{-\varepsilon_{m}^{2}u}}{
  \varepsilon_{m}^{2} -\varepsilon_{n}^{2}}-\frac{\varepsilon_{n}
  e^{-u\varepsilon_{n}^{2}}
  +\varepsilon_{m}e^{-u\varepsilon_{m}^{2}}}{ \varepsilon_{m} +
  \varepsilon_{n}}\Big],\cr
\mathcal{R}_{4}(\varepsilon_{n},\varepsilon_{m}) & = & \frac{1}{2\pi}
\int_{1/\Lambda^{2}}^{\infty} du\int_{0}^{1}d\alpha
e^{-\varepsilon_{n}^{2}u(1-\alpha) -\alpha\varepsilon_{m}^{2}u}
\frac{\varepsilon_{n} (1-\alpha)-\alpha\varepsilon_{m}}{
  \sqrt{\alpha(1-\alpha)}}, \cr
\mathcal{R}_{5}(\varepsilon_{n},\varepsilon_{m}) & = & \frac{1}{2}
\frac{\textrm{sign} \varepsilon_{n}-\textrm{sign} \varepsilon_{m}}{
  \varepsilon_{n} -\varepsilon_{m}},\cr
\mathcal{R}_{6}(\varepsilon_{n},\varepsilon_{m}) & = &
\frac{1-\textrm{sign}(\varepsilon_{n})\textrm{sign}
(\varepsilon_{m})}{\varepsilon_{n}  -\varepsilon_{m}}.
\end{eqnarray}

\section{Baryon matrix elements \label{CGs}}

For the magnetic and axial-vector constants, we can write
Eqs.(\ref{eq:Mag dens},\ref{eq:Ax dens})
in the following forms:
\begin{eqnarray}
\mathcal{G}_{M}^{\chi}(0) & = &
w_{1}D_{\chi3}^{(8)}+w_{2}d_{pq3}D_{\chi
  p}^{(8)}\hat{J_{q}}+w_{3}\frac{1}{\sqrt{3}}
D_{\chi8}^{(8)}\hat{J}_{3}\nonumber \\
 &  & +w_{4}\frac{1}{\sqrt{3}}d_{pq3}D_{\chi
   p}^{(8)}D_{8q}^{(8)}+w_{5}(D_{\chi3}^{(8)}D_{88}^{(8)} +
 D_{\chi8}^{(8)}D_{83}^{(8)})  \nonumber  \\
 &  & +w_{6}(D_{\chi3}^{(8)}D_{88}^{(8)} -
 D_{\chi8}^{(8)}D_{83}^{(8)}),
\label{GMmodel}
\end{eqnarray}
and
\begin{eqnarray}
\mathcal{G}_{10}^{\chi}(0) & = &
a_{1}D_{\chi3}^{(8)}+a_{2}d_{pq3}D_{\chi
  p}^{(8)}J_{q}+\frac{a_{3}}{\sqrt{3}}\,D_{\chi8}^{(8)}J_{3} \cr
 &  & +\frac{a_{4}}{\sqrt{3}}\, d_{pq3}D_{\chi
   p}^{(8)}D_{8q}^{(8)}+a_{5}\Big[D_{\chi3}^{(8)}D_{88}^{(8)} +
 D_{\chi8}^{(8)}D_{83}^{(8)}\Big] \cr
 &  &
 +a_{6}\Big[D_{\chi3}^{(8)}D_{88}^{(8)}-D_{\chi8}^{(8)}D_{83}^{(8)}\Big].
 \label{eq:GA in a1,a2..}
\end{eqnarray}
The above densities were evaluated for the constituent quark mass
$M=420$ MeV and box size of $8$ fm and yield the parameters $w_i$,
$a_i$ as: Previous numbers for $w_i$ were from the normalization of
magnetic moments with the experimental nucleon mass. These numbers are
now with normalization to the soliton-nucleon mass. 
\begin{center}
\begin{tabular}{cccccc}
\hline
$w_{1}$&
$w_{2}$&
$w_{3}$&
$w_{4}$&
$w_{5}$&
$w_{6}$\tabularnewline
\hline
$\begin{array}{c}
-12.94\,(\textrm{with}\,\mathcal{M}_{0})\\
-13.64\,(\textrm{without}\,\mathcal{M}_{0})\end{array}$&
$7.13$&
$5.16$&
$-1.31$&
$-0.78$&
$0.07$\tabularnewline
\hline
\end{tabular}
\end{center}

\begin{center}
\begin{tabular}{cccccc}
\hline
$a_{1}$&
$a_{2}$&
$a_{3}$&
$a_{4}$&
$a_{5}$&
$a_{6}$\tabularnewline
\hline
$\begin{array}{c}
-3.70\,(\textrm{with}\,\mathcal{H})\\
-3.64\,(\textrm{without}\,\mathcal{H})\end{array}$&
$2.50$&
$0.90$&
$-0.18$&
$0.02$&
$0.04$\tabularnewline
\hline
\end{tabular}

\end{center}
where the parameters $w_1$ and $a_1$ contain the $m_s$ corrections due
to $\mathcal{M}_0$ and $\mathcal{H}$.  All magnetic and axial-vector
constants in this work can be reproduced by these parameters.

We list now the results of the matrix elements needed for the
electric, magnetic and axial-vector form factors, where we make the
following abbreviations:
\textbf{\[
d_{ab3}D_{3b}^{(8)}J_{a}=dD_{3}J,\,\,\,\,\,\,\,\,\,\,
d_{ab3}D_{8b}^{(8)}J_{a}=dD_{8}J,\,\,\,\,\,\,\,\,\,\, S_{3}=1/2\]
}
The following matrix elements for the axial-vector and magnetic form
factors are consistent with those given in
Ref.~\cite{semilep_modelindep}. The operators are
$\chi=\sqrt{2}\Sigma^-$ and $\chi = \sqrt{2}\Xi^-$ for the $\Delta
S=0$ and $\Delta S=1$ transitions, respectively. The isospin relations
are:
\begin{eqnarray}
\Sigma_{8}^{-}\to\Lambda_{8}&=&-\Big[\Sigma_{8}^{+}\to\Lambda_{8}\Big],\;\;\;
\Sigma_{8}^{-}\to\Sigma_{8}^{0}=\Sigma_{8}^{0}\to\Sigma_{8}^{+},\cr
\Sigma_{8}^{-}\to n_{8}&=&\sqrt{2}\Big[\Sigma_{8}^{0}\to p_{8}\Big],\;\;\;
\Xi_{8}^{-}\to\Sigma_{8}^{0}=\frac{1}{\sqrt{2}}
\Big[\Xi_{8}^{0}\to\Sigma_{8}^{+}\Big].
\end{eqnarray}

Note that the overall factor of $1/2$ in the definition of the quark
current operator is not included in the following matrix elements. 
\vspace{0.3cm}

\begin{tabular}{c|c|c|c}
\hline
{ $n\to p$ }&
{ $\langle p_{8}|D|n_{8}\rangle$}&
&
{ $\langle p_{8}|D|n_{8}\rangle$}\tabularnewline
\hline
{ $D_{\sqrt{2}\Sigma^{-}3}^{8}$}&
{ $-\frac{7}{15}-c_{\overline{10}}\frac{2}{3}-c_{27}\frac{4}{45}$}&
{ $D_{88}^{(8)}D_{\sqrt{2}\Sigma^{-}3}^{(8)}$}&
{ $-\frac{8}{45}$}\tabularnewline
{ $D_{\sqrt{2}\Sigma^{-}8}^{(8)}J_{3}$}&
{ $\frac{1}{2}\Big[\frac{1}{5}\sqrt{\frac{1}{3}}-c_{\overline{10}}2
\sqrt{\frac{1}{3}}+c_{27}\frac{4}{5}\sqrt{\frac{1}{3}}\Big]$}& 
{ $D_{83}^{(8)}D_{\sqrt{2}\Sigma^{-}8}^{(8)}$}&
{ $-\frac{2}{45}$}\tabularnewline
{ $d_{ab3}D_{\sqrt{2}\Sigma^{-}b}^{(8)}J_{a}$}&
{ $\frac{7}{30}-c_{\overline{10}}\frac{2}{3}-c_{27}\frac{8}{45}$}&
{ $D_{8a}^{(8)}D_{\sqrt{2}\Sigma^{-}b}^{(8)}d_{ab3}$}&
{ $-\frac{11}{45}\sqrt{\frac{1}{3}}$}\tabularnewline
\hline
\end{tabular}{ \par}

{ }\begin{tabular}{c|c|c|c}
\hline
{ $\Sigma_{8}^{-}\to\Lambda_{8}$ }&
{ $\langle\Lambda_{8}\mid D\mid\Sigma_{8}^{-}\rangle$}&
&
{ $\langle\Lambda_{8}\mid D\mid\Sigma_{8}^{-}\rangle$}\tabularnewline
\hline
{ $D_{\sqrt{2}\Sigma^{-}3}^{8}$}&
{ $-\frac{3}{5}\sqrt{\frac{1}{6}}-c_{\overline{10}}\sqrt{\frac{1}{6}}
  - c_{27}\frac{1}{3}\sqrt{\frac{1}{6}}$}& 
{ $D_{88}^{(8)}D_{\sqrt{2}\Sigma^{-}3}^{(8)}$}&
{ $-\frac{1}{15}\sqrt{\frac{1}{6}}$}\tabularnewline
{ $D_{\sqrt{2}\Sigma^{-}8}^{(8)}J_{3}$}&
{ $\frac{1}{2}\Big[-\frac{1}{5}\sqrt{\frac{1}{2}}-c_{\bar{10}}
  \sqrt{\frac{1}{2}} +c_{27}\sqrt{\frac{1}{2}}\Big]$}&
{ $D_{83}^{(8)}D_{\sqrt{2}\Sigma^{-}8}^{(8)}$}&
{ $-\frac{1}{15}\sqrt{\frac{1}{6}}$}\tabularnewline
{ $d_{ab3}D_{\sqrt{2}\Sigma^{-}b}^{(8)}J_{a}$}&
{ $\frac{3}{10}\sqrt{\frac{1}{6}}-c_{\bar{10}} \sqrt{\frac{1}{6}} -
  c_{27}\frac{2}{3}\sqrt{\frac{1}{6}}$}& 
{ $D_{8a}^{(8)}D_{\sqrt{2}\Sigma^{-}b}^{(8)}d_{ab3}$}&
{ $-\frac{7}{45}\sqrt{\frac{1}{2}}$}\tabularnewline
\hline
\end{tabular}{ \par}

{ }\begin{tabular}{c|c|c|c}
\hline
{ $\Sigma_{8}^{-}\to\Sigma_{8}^{0}$}&
{ $\langle\Sigma_{8}^{0}\mid D\mid\Sigma_{8}^{-}\rangle$}&
&
{ $\langle\Sigma_{8}^{0}\mid D\mid\Sigma_{8}^{-}\rangle$}\tabularnewline
\hline
{ $D_{\sqrt{2}\Sigma^{-}3}^{8}$}&
{ $-\frac{1}{3}\sqrt{\frac{1}{2}}-c_{\bar{10}}\frac{2}{3}\sqrt{\frac{1}{2}}$}&
{ $D_{88}^{(8)}D_{\sqrt{2}\Sigma^{-}3}^{(8)}$}&
{ $0$}\tabularnewline
{ $D_{\sqrt{2}\Sigma^{-}8}^{(8)}J_{3}$}&
{ $\frac{1}{2}\Big[\sqrt{\frac{1}{6}}-c_{\overline{10}}2\sqrt{\frac{1}{6}}\Big]$}&
{ $D_{83}^{(8)}D_{\sqrt{2}\Sigma^{-}8}^{(8)}$}&
{ $-\frac{2}{15}\sqrt{\frac{1}{2}}$}\tabularnewline
{ $d_{ab3}D_{\sqrt{2}\Sigma^{-}b}^{(8)}J_{a}$}&
{ $\frac{1}{6}\sqrt{\frac{1}{2}}-c_{\bar{10}}\frac{2}{3}\sqrt{\frac{1}{2}}$}&
{ $D_{8a}^{(8)}D_{\sqrt{2}\Sigma^{-}b}^{(8)}d_{ab3}$}&
{ $-\frac{1}{5}\sqrt{\frac{1}{6}}$}\tabularnewline
\hline
\end{tabular}{ \par}

{ }\begin{tabular}{c|c|c|c}
\hline
{ $\Xi_{8}^{-}\to\Xi_{8}^{0}$}&
{ $\langle\Xi_{8}^{0}\mid D\mid\Xi_{8}^{-}\rangle$}&
&
{ $\langle\Xi_{8}^{0}\mid D\mid\Xi_{8}^{-}\rangle$}\tabularnewline
\hline
{ $D_{\sqrt{2}\Sigma^{-}3}^{8}$}&
{ $\frac{2}{15}+c_{27}\frac{4}{45}$}&
{ $D_{88}^{(8)}D_{\sqrt{2}\Sigma^{-}3}^{(8)}$}&
{ $-\frac{1}{45}$}\tabularnewline
{ $D_{\sqrt{2}\Sigma^{-}8}^{(8)}J_{3}$}&
{ $\frac{1}{2}\Big[\frac{4}{5}\sqrt{\frac{1}{3}} - c_{27}\frac{4}{5}
  \sqrt{\frac{1}{3}}\Big]$}& 
{ $D_{83}^{(8)}D_{\sqrt{2}\Sigma^{-}8}^{(8)}$}&
{ $\frac{1}{9}$}\tabularnewline
{ $d_{ab3}D_{\sqrt{2}\Sigma^{-}b}^{(8)}J_{a}$}&
{ $-\frac{1}{15}+c_{27}\frac{8}{45}$}&
{ $D_{8a}^{(8)}D_{\sqrt{2}\Sigma^{-}b}^{(8)}d_{ab3}$}&
{ $\frac{2}{45}\sqrt{\frac{1}{3}}$}\tabularnewline
\hline
\end{tabular}

\begin{tabular}{c|c|c|c}
\hline
{ $\Sigma_{8}^{-}\to n_{8}$}&
{ $\langle n_{8}\mid D\mid\Sigma_{8}^{-}\rangle$}&
&
{ $\langle n_{8}\mid D\mid\Sigma_{8}^{-}\rangle$}\tabularnewline
\hline
{ $D_{\sqrt{2}\Xi^{-}3}^{8}$}&
{ $\frac{2}{15}-c_{27}\frac{2}{45}$}&
{ $D_{88}^{(8)}D_{\sqrt{2}\Xi^{-}3}^{(8)}$}&
{ $\frac{1}{90}$}\tabularnewline
{ $D_{\sqrt{2}\Xi^{-}8}^{(8)}J_{3}$}&
{ $\frac{1}{2}\Big[\frac{4}{5}\sqrt{\frac{1}{3}} +
  c_{27}\frac{2}{5}\sqrt{\frac{1}{3}}\Big]$}& 
{ $D_{83}^{(8)}D_{\sqrt{2}\Xi^{-}8}^{(8)}$}&
{ $-\frac{1}{18}$}\tabularnewline
{ $d_{ab3}D_{\sqrt{2}\Xi^{-}b}^{(8)}J_{a}$}&
{ $-\frac{1}{15}-c_{27}\frac{4}{45}$}&
{ $D_{8a}^{(8)}D_{\sqrt{2}\Xi^{-}b}^{(8)}d_{ab3}$}&
{ $-\frac{1}{45}\sqrt{\frac{1}{3}}$}\tabularnewline
\hline
\end{tabular}{ \par}

{ }\begin{tabular}{c|c|c|c}
\hline
{ $\Lambda_{8}\to p_{8}$}&
{ $\langle p_{8}\mid D\mid\Lambda_{8}\rangle$}&
&
{ $\langle p_{8}\mid D\mid\Lambda_{8}\rangle$}\tabularnewline
\hline
{ $D_{\sqrt{2}\Xi^{-}3}^{8}$}&
{ $-\frac{4}{5}\sqrt{\frac{1}{6}} +
  c_{\overline{10}}\sqrt{\frac{1}{6}} +
  c_{27}\frac{1}{5}\sqrt{\frac{1}{6}}$}& 
{ $D_{88}^{(8)}D_{\sqrt{2}\Xi^{-}3}^{(8)}$}&
{ $-\frac{1}{10}\sqrt{\frac{1}{6}}$}\tabularnewline
{ $D_{\sqrt{2}\Xi^{-}8}^{(8)}J_{3}$}&
{ $\frac{1}{2}\Big[\frac{2}{5}\sqrt{\frac{1}{2}} + c_{\overline{10}}
  \sqrt{\frac{1}{2}} -c_{27}\frac{3}{5}\sqrt{\frac{1}{2}}\Big]$}& 
{ $D_{83}^{(8)}D_{\sqrt{2}\Xi^{-}8}^{(8)}$}&
{ $\frac{1}{10}\sqrt{\frac{1}{6}}$}\tabularnewline
{ $d_{ab3}D_{\sqrt{2}\Xi^{-}b}^{(8)}J_{a}$}&
{ $\frac{2}{5}\sqrt{\frac{1}{6}}+c_{\overline{10}}\sqrt{\frac{1}{6}} +
  c_{27}\frac{2}{5}\sqrt{\frac{1}{6}}$}&
{ $D_{8a}^{(8)}D_{\sqrt{2}\Xi^{-}b}^{(8)}d_{ab3}$}&
{ $\frac{2}{15}\sqrt{\frac{1}{2}}$}\tabularnewline
\hline
\end{tabular}{ \par}

{ }\begin{tabular}{c|c|c|c}
\hline
{ $\Xi_{8}^{-}\to\Sigma_{8}^{0}$}&
{ $\langle\Sigma_{8}^{0}\mid D\mid\Xi_{8}^{-}\rangle$}&
&
{ $\langle\Sigma_{8}^{0}\mid D\mid\Xi_{8}^{-}\rangle$}\tabularnewline
\hline
{ $D_{\sqrt{2}\Xi^{-}3}^{8}$}&
{ $-\frac{7}{15}\sqrt{\frac{1}{2}}+c_{\overline{10}}\frac{1}{3}
  \sqrt{\frac{1}{2}} +c_{27}\frac{2}{45}\sqrt{\frac{1}{2}}$}&
{ $D_{88}^{(8)}D_{\sqrt{2}\Xi^{-}3}^{(8)}$}&
{ $\frac{4}{45}\sqrt{\frac{1}{2}}$}\tabularnewline
{ $D_{\sqrt{2}\Xi^{-}8}^{(8)}J_{3}$}&
{ $\frac{1}{2}\Big[\frac{1}{5}\sqrt{\frac{1}{6}} +
  c_{\overline{10}}\sqrt{\frac{1}{6}} - c_{27}\frac{2}{5}
  \sqrt{\frac{1}{6}}\Big]$}& 
{ $D_{83}^{(8)}D_{\sqrt{2}\Xi^{-}8}^{(8)}$}&
{ $\frac{1}{45}\sqrt{\frac{1}{2}}$}\tabularnewline
{ $d_{ab3}D_{\sqrt{2}\Xi^{-}b}^{(8)}J_{a}$}&
{ $\frac{7}{30}\sqrt{\frac{1}{2}}+c_{\overline{10}}
  \frac{1}{3}\sqrt{\frac{1}{2}} +
  c_{27}\frac{4}{45}\sqrt{\frac{1}{2}}$}& 
{ $D_{8a}^{(8)}D_{\sqrt{2}\Xi^{-}b}^{(8)}d_{ab3}$}&
{ $\frac{11}{90}\sqrt{\frac{1}{6}}$}\tabularnewline
\hline
\end{tabular}{ \par}

{ }\begin{tabular}{c|c|c|c}
\hline
{ $\Xi_{8}^{-}\to\Lambda_{8}$ }&
{ $\langle\Lambda_{8}\mid D\mid\Xi_{8}^{-}\rangle$}&
&
{ $\langle\Lambda_{8}\mid D\mid\Xi_{8}^{-}\rangle$}\tabularnewline
\hline
{ $D_{\sqrt{2}\Xi^{-}3}^{8}$}&
{ $-\frac{1}{5}\sqrt{\frac{1}{6}}-c_{27}\frac{1}{5}\sqrt{\frac{1}{6}}$}&
{ $D_{88}^{(8)}D_{\sqrt{2}\Xi^{-}3}^{(8)}$}&
{ $0$}\tabularnewline
{ $D_{\sqrt{2}\Xi^{-}8}^{(8)}J_{3}$}&
{ $\frac{1}{2}\Big[\frac{3}{5}\sqrt{\frac{1}{2}} +
  c_{27}\frac{3}{5}\sqrt{\frac{1}{2}}\Big]$}& 
{ $D_{83}^{(8)}D_{\sqrt{2}\Xi^{-}8}^{(8)}$}&
{ $\frac{1}{5}\sqrt{\frac{1}{6}}$}\tabularnewline
{ $d_{ab3}D_{\sqrt{2}\Xi^{-}b}^{(8)}J_{a}$}&
{ $\frac{1}{10}\sqrt{\frac{1}{6}}-c_{27}\frac{2}{5}\sqrt{\frac{1}{6}}$}&
{ $D_{8a}^{(8)}D_{\sqrt{2}\Xi^{-}b}^{(8)}d_{ab3}$}&
{ $-\frac{1}{30}\sqrt{\frac{1}{2}}$}\tabularnewline
\hline
\end{tabular}{ }\\
{ \par}
\vspace{0.5cm}

Matrix-elements needed for the electric form factor: \\

{\scriptsize }\begin{tabular}{c|ccc}
\hline
&
{\scriptsize $D_{\chi8}^{(8)}$}&
{\scriptsize $D_{\chi i}^{(8)}J_{i}$}&
{\scriptsize $D_{\chi a}^{(8)}J_{a}$}\tabularnewline
\hline
{\scriptsize $\langle p|\Big[\chi=\sqrt{2}\Sigma^{-}\Big]|n\rangle$}&
{\scriptsize
  $+\frac{1}{5}\sqrt{\frac{1}{3}}-c_{\bar{10}}2\sqrt{\frac{1}{3}} +
  c_{27}\frac{4}{5}\sqrt{\frac{1}{3}}$}& 
{\scriptsize $-\frac{7}{10}-c_{\bar{10}}-c_{27}\frac{2}{15}$}&
{\scriptsize $-\frac{1}{5}+c_{27}\frac{8}{15}$}\tabularnewline
{\scriptsize $\langle\Lambda|\Big[\chi=\sqrt{2}\Sigma^{-}\Big]|
  \Sigma^{-}\rangle$}& 
{\scriptsize $-\frac{1}{5}\sqrt{\frac{1}{2}} -
  c_{\bar{10}}\sqrt{\frac{1}{2}} +c_{27}\sqrt{\frac{1}{2}}$}&
{\scriptsize $-\frac{9}{10}\sqrt{\frac{1}{6}} -
  c_{\bar{10}}\frac{3}{2} \sqrt{\frac{1}{6}} -
  c_{27}\frac{1}{2}\sqrt{\frac{1}{6}}$}& 
{\scriptsize $+\frac{3}{5}\sqrt{\frac{1}{6}} +
  c_{27}2\sqrt{\frac{1}{6}}$}\tabularnewline 
{\scriptsize $\langle\Sigma^{0}|\Big[\chi = \sqrt{2}\Sigma^{-}\Big]|
  \Sigma^{-}\rangle$}& 
{\scriptsize $+\sqrt{\frac{1}{6}}-c_{\bar{10}}2\sqrt{\frac{1}{6}}$}&
{\scriptsize
  $-\frac{1}{2}\frac{1}{\sqrt{2}}-c_{\bar{10}}\sqrt{\frac{1}{2}}$}& 
{\scriptsize $-\frac{1}{\sqrt{2}}$}\tabularnewline
{\scriptsize
  $\langle\Xi^{0}|\Big[\chi=\sqrt{2}\Sigma^{-}\Big]|\Xi^{-}\rangle$}& 
{\scriptsize
  $+\frac{4}{5}\sqrt{\frac{1}{3}}-c_{27}\frac{4}{5}
  \sqrt{\frac{1}{3}}$}&  
{\scriptsize $+\frac{1}{5}+c_{27}\frac{2}{15}$}&
{\scriptsize $-\frac{4}{5}-c_{27}\frac{8}{15}$}\tabularnewline
\hline
{\scriptsize $\langle n|\Big[\chi=\sqrt{2}\Xi^{-}\Big]|
  \Sigma^{-}\rangle$}& 
{\scriptsize $+\frac{4}{5}\sqrt{\frac{1}{3}}+c_{27}\frac{2}{5}
  \sqrt{\frac{1}{3}}$}& 
{\scriptsize $+\frac{1}{5}-c_{27}\frac{1}{15}$}&
{\scriptsize $-\frac{4}{5}+c_{27}\frac{4}{15}$}\tabularnewline
{\scriptsize $\langle p|\Big[\chi=\sqrt{2}\Xi^{-}\Big]| \Lambda\rangle$}&
{\scriptsize
  $+\frac{2}{5}\frac{1}{\sqrt{2}}+c_{\bar{10}}\sqrt{\frac{1}{2}} -
  c_{27}\frac{3}{5}\sqrt{\frac{1}{2}}$}& 
{\scriptsize $-\frac{6}{5}\sqrt{\frac{1}{6}} + c_{\bar{10}}\frac{3}{2}
  \sqrt{\frac{1}{6}}+c_{27}\frac{3}{10}\sqrt{\frac{1}{6}}$}&
{\scriptsize $-\frac{6}{5}\sqrt{\frac{1}{6}}-c_{27}\frac{6}{5}
  \sqrt{\frac{1}{6}}$}\tabularnewline 
{\scriptsize $\langle\Sigma^{0}|\Big[\chi=\sqrt{2}\Xi^{-}\Big]|
  \Xi^{-}\rangle$}& 
{\scriptsize $+\frac{1}{5}\sqrt{\frac{1}{6}} +
  c_{\bar{10}}\sqrt{\frac{1}{6}} -c_{27}\frac{6}{15}\sqrt{\frac{1}{6}}$}&
{\scriptsize $-\frac{7}{10}\sqrt{\frac{1}{2}} +
  c_{\bar{10}}\frac{1}{2} \sqrt{\frac{1}{2}} +
  c_{27}\frac{1}{15}\sqrt{\frac{1}{2}}$}& 
{\scriptsize $-\frac{1}{5}\sqrt{\frac{1}{2}} -
  c_{27}\frac{4}{15}\sqrt{\frac{1}{2}}$} \tabularnewline
{\scriptsize $\langle\Lambda|\Big[\chi =
  \sqrt{2}\Xi^{-}\Big]|\Xi^{-}\rangle$}& 
{\scriptsize $+\frac{3}{5}\sqrt{\frac{1}{2}} +
  c_{27}\frac{3}{5}\sqrt{\frac{1}{2}}$}& 
{\scriptsize $-\frac{3}{10}\sqrt{\frac{1}{6}} -
  c_{27}\frac{3}{10}\sqrt{\frac{1}{6}}$}& 
{\scriptsize $-\frac{9}{5}\sqrt{\frac{1}{6}} +
  c_{27}\frac{6}{5}\sqrt{\frac{1}{6}}$}\tabularnewline 
\hline
\end{tabular}{\scriptsize \par}

{  }\begin{tabular}{c|ccc}
\hline
&
{  $D_{88}^{(8)}D_{\chi8}^{(8)}$}&
{  $D_{8i}^{(8)}D_{\chi i}^{(8)}$}&
{  $D_{8a}^{(8)}D_{\chi a}^{(8)}$}\tabularnewline
\hline
{  $\langle p|\Big[\chi=\sqrt{2}\Sigma^{-}\Big]|n\rangle$}&
{  $0$}&
{  $+\frac{2}{15}\sqrt{\frac{1}{3}}$}&
{  $-\frac{2}{15}\sqrt{\frac{1}{3}}$}\tabularnewline
{  $\langle\Lambda|\Big[\chi=\sqrt{2}\Sigma^{-}\Big]|\Sigma^{-}\rangle$}&
{  $0$}&
{  $-\frac{2}{15}\sqrt{\frac{1}{2}}$}&
{  $+\frac{2}{15}\sqrt{\frac{1}{2}}$}\tabularnewline
{  $\langle\Sigma^{0}|\Big[\chi=\sqrt{2}\Sigma^{-}\Big]|\Sigma^{-}\rangle$}&
{  $-\frac{1}{5}\sqrt{\frac{1}{6}}$}&
{  $+\frac{3}{5}\sqrt{\frac{1}{6}}$}&
{  $-\frac{2}{5}\sqrt{\frac{1}{6}}$}\tabularnewline
{  $\langle\Xi^{0}|\Big[\chi=\sqrt{2}\Sigma^{-}\Big]|\Xi^{-}\rangle$}&
{  $-\frac{1}{5}\sqrt{\frac{1}{3}}$}&
{  $+\frac{7}{15}\sqrt{\frac{1}{3}}$}&
{  $-\frac{4}{15}\sqrt{\frac{1}{3}}$}\tabularnewline
\hline
{  $\langle n|\Big[\chi=\sqrt{2}\Xi^{-}\Big]|\Sigma^{-}\rangle$}&
{  $+\frac{1}{10}\sqrt{\frac{1}{3}}$}&
{  $-\frac{7}{30}\sqrt{\frac{1}{3}}$}&
{  $+\frac{2}{15}\sqrt{\frac{1}{3}}$}\tabularnewline
{  $\langle p|\Big[\chi=\sqrt{2}\Xi^{-}\Big]|\Lambda\rangle$}&
{  $+\frac{1}{10}\sqrt{\frac{1}{2}}$}&
{  $-\frac{1}{10}\sqrt{\frac{1}{2}}$}&
{  $0$}\tabularnewline
{  $\langle\Sigma^{0}|\Big[\chi=\sqrt{2}\Xi^{-}\Big]|\Xi^{-}\rangle$}&
{  $0$}&
{  $-\frac{1}{15}\sqrt{\frac{1}{6}}$}&
{  $+\frac{1}{15}\sqrt{\frac{1}{6}}$}\tabularnewline
{  $\langle\Lambda|\Big[\chi=\sqrt{2}\Xi^{-}\Big]|\Xi^{-}\rangle$}&
{  $0$}&
{  $-\frac{1}{5}\sqrt{\frac{1}{2}}$}&
{  $+\frac{1}{5}\sqrt{\frac{1}{2}}$}\tabularnewline
\hline
\end{tabular}{  \par}
\end{appendix}

\end{document}